\documentclass[nofootinbib]{article}

\usepackage{color}

\usepackage{amsfonts, amsmath, graphicx, psfrag, array, bbm, float, multirow}

\usepackage{color}

\usepackage{hyperref}

\definecolor{linkc}{rgb}{0,0,0.3}
\definecolor{rule}{rgb}{0.7,0.7,0.7}

\hypersetup{colorlinks=true, linkcolor=black, urlcolor=linkc, citecolor=linkc, bookmarksopen}

\newcommand{\dsty}{\displaystyle}

\newcommand{\be}{\begin{equation}}
\newcommand{\ee}{\end{equation}}

\newcommand{\bes}{\begin{eqnarray}}
\newcommand{\ees}{\end{eqnarray}}

\newcommand{\ba}{\begin{array}}
\newcommand{\ea}{\end{array}}

\newcommand{\so}{\mathfrak{so}}

\newcommand{\su}{\mathfrak{su}}

\newcommand{\SO}{\textrm{SO}}

\newcommand{\SU}{\textrm{SU}}

\newcommand{\R}{\mathbb{R}}

\newcommand{\cA}{\mathcal{A}}
\newcommand{\cB}{\mathcal{B}}
\newcommand{\cC}{\mathcal{C}}
\newcommand{\cD}{\mathcal{D}}
\newcommand{\cE}{\mathcal{E}}

\newcommand{\cG}{\mathcal{G}}

\newcommand{\cM}{\mathcal{M}}

\newcommand{\cP}{\mathcal{P}}

\newcommand{\cS}{\mathcal{S}}

\newcommand{\sca}{\textsc{a}}

\newcommand{\scc}{\textsc{c}}
\newcommand{\scd}{\textsc{d}}
\newcommand{\sce}{\textsc{e}}
\newcommand{\scf}{\textsc{f}}
\newcommand{\scg}{\textsc{g}}

\newcommand{\scm}{\textsc{m}}

\newcommand{\sct}{\textsc{t}}

\newcommand{\dl}{\stackrel{L}{\Delta}}
\newcommand{\dli}{\stackrel{L}{\Delta}\stackrel{}{{}^{-1}}}
\newcommand{\dr}{\stackrel{R}{\Delta}\;}
\newcommand{\dri}{\stackrel{R}{\Delta}\stackrel{}{{}^{-1}}}

\newcommand{\cl}{\stackrel{L}{\mathcal{S}}}

\newcommand{\ua}{{}^{(1)}}
\newcommand{\ub}{{}^{(2)}}

\newcommand{\cspacetime}{\mathcal{M}}
\newcommand{\cspace}{\Sigma}

\newcommand{\dspace}{\Delta_3}

\newcommand{\covd}{\mathcal{D}}

\newcommand{\g}{\gamma}

\begin{document}



\title{\bf On the role of the Barbero-Immirzi parameter in discrete quantum gravity}

\author{
 Bianca Dittrich$^{1,2}$, James P.~Ryan$^2$\\[0.2cm]
\small $^1$ Perimeter Institute, 
\small  31 Caroline St. N, Waterloo, ON N2L 2Y5, Canada\\ 
\small $^2$ MPI f\"ur Gravitationsphysik, Am M\"uhlenberg 1, D-14476 Potsdam, Germany
 }

\date{}

\maketitle

\begin{abstract}
\noindent  The 1-parameter family of transformations identified by Barbero and Immirzi plays a significant role in non-perturbative approaches to quantum gravity, among them Loop Quantum Gravity and Spin Foams. It facilitates the loop quantization programme and subsequently the Barbero-Immirzi parameter ($\gamma$) arises in both the spectra of geometrical operators and in the dynamics provided by Spin Foams. However, the debate continues as to whether quantum physics should be Barbero-Immirzi parameter dependent. 
Starting from a discrete $\SO(4)$-$BF$ theory phase space, we find two possible reductions with respect to a discrete form of the simplicity constraints.  The first reduces to a phase space with $\gamma$-dependent symplectic structure and more generally in agreement with the phase space underlying Loop Quantum Gravity restricted to a single graph - a.k.a.\ Twisted Geometries.  The second, fuller reduction leads to a $\gamma$-independent symplectic structure on the phase space of piecewise-flat-linear geometries - a.k.a Regge geometries. Thus, the $\gamma$-dependence of physical predictions is related to the choice of phase space underlying the quantization.

\end{abstract}




\section{Introduction}

Since its inception, the study of the gravitational force has proved to be a fruitful yet frequently perplexing endeavour.  With the advent of quantum theory and its manifold applications to the description of matter, the quantization of gravity seems a natural step towards a more cohesive picture of the world around us. Holding true to form, however, the twin concepts of dynamical geometry and diffeomorphism invariance appear to lie at the root of gravity's stubborn resistance to a fundamental quantum mechanical description.  

For quite some time now, there have existed a number of approaches to this challenge that, in one manner or another, involve replacing the continuum manifold by a simplicial one.  We shall mention just two of these: quantum Regge calculus \cite{Rocek:1982fr,Rocek:1982tj, Regge:2000wu} and the more modern Spin Foam models \cite{Engle:2007wy, Freidel:2007py}. The former is a Lagrangian path integral approach where all the paths have support on a fixed simplicial manifold and 
are weighted by the exponential of the Regge action - a discretization of the Einstein-Hilbert action.  The latter are state sum models that can be interpreted in various ways (a property that makes them all the more appealing) and using different kind of variables, for instance  non-commutative first order variables \cite{Baratin:2011tx,Baratin:2011hp}, twisted geometry variables \cite{Speziale:2012np} or just holonomies \cite{Bahr:2012qj, Dittrich:2012er}. Spin foam models can also be viewed as a Lagrangian path integral approach \cite{Bonzom:2008ru,Bonzom:2009wm,Bonzom:2009hw}, where once again all the paths have support on a fixed simplicial (or even cellular) manifold and are weighted by the exponential of a discretized form of the Holst-modified Plebanski action \cite{Holst:1995pc, Plebanski:1977zz}. The presence of the Holst term not only entails that modern spin foam models have a dependence on the Barbero-Immirzi parameter ($\gamma$) \cite{BarberoG.:1993aa,Immirzi:1994gq,Immirzi:1992ar}.   We shall present more details on this a bit later, but for the moment, let us note that quantum Regge calculus has no $\gamma$-dependence.  More importantly, the ultimate significance of this parameter remains unclear.  

\vspace{0.2cm}

Our aim in this paper is to ameliorate this state of affairs by clarifying the role of the Barbero-Immirzi parameter in the classical discrete theory. 

\vspace{0.2cm}

But first, let us begin in the continuum, where a somewhat similar situation exists.  Tersely speaking, Loop Quantum Gravity (LQG) is a traditional canonical quantization procedure applied to gravity that has provided a rigorous definition of the kinematical Hilbert space, along with an anomaly-free action of the Hamiltonian constraint operator \cite{Ashtekar:2004eh, Rovelli:2004tv, Thiemann:2007zz}.  These powerful successes stem from the ability to parameterize the phase space of continuum spatial geometries in terms of an $\SU(2)$ connection and a triad field \cite{Ashtekar:1986yd},  the same variables that parameterize the phase space of $\SU(2)$-$BF$ theory \cite{Birmingham:1991ty}.  Of course, these variable are still subject to further constraints: the rotational Gauss, 3-diffeomorphism and Hamiltonian constraints. The gauge-theoretic profile of this phase space is not a priori apparent, rather it is uncovered via a canonical transformation \cite{BarberoG.:1993aa}.  Remarkably, there is a 1-parameter family of such transformations, labelled by the Barbero-Immirzi parameter.

While the classical theory is, at the end of the day, insensitive to these $\gamma$-dependent transformations, the Barbero-Immirzi parameter subtly finds its way into the quantum theory \cite{Rovelli:1997na}. The point is that transformations with different values of $\gamma$, define different pairs of conjugate (kinematical) observables. These are subsequently quantized, that is, they are represented as operators on a Hilbert space. The quantization procedures involves a choice of holonomies as basic variables, in this case, exponentiated ($\gamma$-dependent) connections. This leads to the Barbero-Immirzi parameter manifesting itself in the Poisson brackets between basic classical variables and therefore in the commutators involving basic quantum operators.\footnote{In particular, let us apply the Jacobi identity, which holds for both Poisson and commutator brackets, to $\{e^{\gamma q},\{p_1,p_2\}\}$; $q, p$ are analogues of the connection and triad, respectively. One notices that  a non--vanishing bracket between momentum variables implies $\{p_1,p_2\}\sim\gamma $, assuming that $\{q,p_i\}= 1$. Indeed, the necessity of non-commuting fluxes in Loop Quantum Gravity \cite{Ashtekar:1998ak,Thiemann:2000bv,Baratin:2010nn} follows from a similar kind of argument.}

This quantization procedure even leads to the appearance of $\gamma$ in the Poisson and commutator brackets between spatial geometric quantities, the 3d-dihedral angles for example, despite $\gamma$ having been introduced only in the connection (which encodes the extrinsic curvature).
 In particular the spectra of geometrical operators, notable the area and volume operators \cite{Rovelli:1994ge, Ashtekar:1996eg, Ashtekar:1997fb}, turn out to depend on $\gamma$ and thus show that quantum theories for different values of $\gamma$ are unitarily inequivalent.

This is not the end of the story in the continuum, however.  One may view the path to this loop gravity phase space within the context of the Holst-modified Einstein-Cartan formulation \cite{Holst:1995pc}. Therein one begins with an auxiliary phase space parameterized by bi-vectors and a Lorentz connection. These variables are identical those of a Lorentzian $BF$ theory and are subject to the famous simplicity constraints (both primary and secondary) and Lorentz Gauss constraint, in addition to the 3-diffeomorphism and Hamiltonian constraints. 

Importantly, the primary and secondary simplicity constraints are 2nd class and therefore in any quantization procedure, they should be dealt with at the classical level.  There are two completely equivalent methods to reduce with respect to these constraints: i) explicit solution or ii) Dirac bracket construction.  

To mention only the pertinent details, explicit solution yields 
a spatial Lorentz connection, whose construction involves $\gamma$, and its conjugate momentum.  Moreover, a subsequent gauge-fixing of the boost part of the Gauss constraint lands us back in the phase space underlying the LQG approach \cite{Geiller:2011cv,Geiller:2011bh}. Thus, one expects that the Barbero-Immirzi parameter still plays a fundamental role after quantization. On the other hand, by constructing a Dirac bracket, one can parameterize the reduced phase space using yet another Lorentz connection (and its conjugate momentum). Remarkably, the Barbero-Immirzi parameter has disappeared, i.e. it does not show up in the Poisson brackets between this connection and conjugated variables nor in the other constraints \cite{Alexandrov:2000jw,Alexandrov:2001wt,Alexandrov:2002br}. Therefore, any subsequent quantization would inherit this property.  Given the presence of the Lorentz connection, it was hoped that one could perform an analogous loop quantization \cite{Alexandrov:2002xc}. Unfortunately, this connection is non-commutative (w.r.t.\ the Dirac bracket) and as a result, only a formal quantization has been achieved. Despite that, it has spurred on much debate as to whether quantum gravitational physics is $\gamma$-dependent \cite{deBerredoPeixoto:2012xd, Alexandrov:2008iy, Mercuri:2006um, Mercuri:2009zt, Benedetti:2011nd,Benedetti:2011yb,Daum:2010qt,Daum:2011bs, Ashtekar:1997yu, Engle:2009vc,Bianchi:2012ui}. 

\vspace{0.5cm}

 As mentioned earlier, the apparent dilemma with respect to $\gamma$-dependence has a counterpart in the discrete setting.  We shall reveal it along the following lines.

Despite the progress in the LQG programme, the analysis of the Hamiltonian constraint operator in the full kinematical Hilbert space has remained for the most part at a rather abstract level; one can formally define a projector onto the physical state space.  As a result, attention has concentrated on providing a definition of the projector onto physical states for spin networks with support on a fixed graph.  In essence, this corresponds to a truncation of the theory, with some very interesting repercussions. One considers a spin-network as the dual graph to a cellular decomposition of the spatial 3-manifold. Furthermore, there is a classical phase space underlying that part of the Hilbert space associated to each spin network graph.  As one might imagine from our earlier discussion, it is essentially the phase space of discrete $\SU(2)$-$BF$ theory.  However, it may be recast in a more geometrical fashion known as Twisted geometries \cite{Freidel:2010aq}
and is a discretization of the phase space underlying LQG.  Thus, as we shall detail later on, the Poisson brackets between basic variables, and the constraints have explicit Barbero-Immirzi parameter dependence, leading again to the appearance of $\gamma$ in the spectrum of geometric operators.
Consequently, the standard spin foam quantization procedure results in $\gamma$-dependent amplitudes weighting the paths.

However, there is another method to quantize gravity in the discrete setting known as quantum Regge calculus, which is $\gamma$-independent.  Recently, a phase space has been prescribed for the classical theory \cite{Dittrich:2009fb,Dittrich:2011ke}.  This phase space describes Regge geometries geometries or more precisely, piecewise-linear-flat geometries.  

It should be noted that Twisted geometries are not identical to Regge geometries. Regge geometries are a more traditional interpretation of discrete geometry in that one has an unambiguous assignment of edges lengths and deficit angles to the simplicial 3-manifold. This is not case for Twisted geometries; while one can construct edge lengths and deficit angles, they are dependent on 
  a choice of tetrahedron and edge respectively, whose data are used to construct these geometrical quantities.

As a consequence, an important statement can be made at this point. In the continuum, the variables of $\SU(2)$-$BF$ theory capture spatial geometries, as well as the conjugate extrinsic curvatures. In the discrete, they capture a generalized class of spatial discrete geometries (and conjugate variables).  The discrete $SU(2)$-BF phase space, as well as the continuum one, lead to $\gamma$-dependent Poisson brackets between basic variables and consequently to $\gamma$-dependent spectra of geometric quantum observables. This opens the question to whether a further reduction of the discrete phase space to a phase space describing Regge geometries is possible and if so, whether such a description would involve $\gamma$ in any way. If the latter were not the case, then $\gamma$ could not be involved in the spectra of any geometrical observables.

We shall see in this paper how the different discrete geometrical phase spaces for are related to each other. More precisely, this paper summarizes relevant findings of two previous papers \cite{Dittrich:2008ar,Dittrich:2010ey} on the various forms of discrete simplicity constraints and gives the analysis of the various Dirac bracket structures, so as to infer the status of $\gamma$ in the different phase spaces.   In Section \ref{sec:plebanski}, we give a very brief description of the continuum Holst phase space.  Thereafter, in Section \ref{sec:poisson},  we discretize the spatial manifold along with the dynamical system. In this canonical setting, this means that we provide a discrete counterpart of the phase space variables, along with an appropriate symplectic structure subject to a discrete version of the constraints. As we shall deal with Riemannian gravity, we shall have at the outset an $\SO(4)$-$BF$ theory phase space subject to a discrete form of the Gauss, simplicity and diffeomorphism constraints.   Our analysis, in Section \ref{sec:reduction},  then concentrates on completing a reduction with respect to the Gauss and simplicity constraints. 

\vspace{0.2cm}

\noindent The upshots of this analysis are the following:

\begin{itemize} 
\item[-] If we reduce by a subset of the discrete simplicity constraints, we arrive at the phase space of Twisted geometries, in its $\gamma$-dependent description.  

\item[-] If we reduce by all the discrete simplicity constraints, we arrive at the phase space of Regge geometries, with $\gamma$-independent Poisson brackets between basic variables.

\end{itemize}

\vspace{0.2cm}

\noindent We can draw a number of conclusions: 

\begin{itemize}
\item[-] In effect, the above results highlight a very common property of the discretization mechanism: at least at the outset, one has a certain amount of freedom when it comes to discretizing a dynamical system.

Here, one such choice is to first reduce the Holst phase space in the continuum down to the loop gravity phase space and then discretize. In this fashion, one is naturally led to the phase space of Twisted geometries. 

Another choice is to first discretize the larger Holst phase space, based on $SO(4)$ holonomies. In this case, one has some freedom in how one discretizes the simplicity constraints.  Of course at this kinematical stage, such choices are based on prejudice. If one is motivated to maintain strict discrete geometricity, one chooses a larger set of discrete constraints and one arrives at the phase space of Regge geometries. 
However, if one is motivated to get to twisted geometries, then one chooses a smaller set of the discrete constraints and one finds  $\gamma$ dependent Poisson brackets between basic variables. Thus the appearance of $\gamma$ can be seen due to an enlargement of the discrete configurations, on which a dynamics is defined. Ultimately, the question will be whether one can find a consistent dynamics based on these configurations, leading to the correct classical limit, in which $\gamma$ should not have any physical significance.


\item[-] 
The respective appearance and disappearance of $\gamma$ in the Twisted and Regge phase spaces and their corresponding quantum theories might not be too surprising. The constraints which would reduce the phase space, equivalent to Twisted geometries, to the Regge phase space are second class. Following standard Dirac procedure, one should solve these constraints classically and perform quantization afterwards.\footnote{Yet another interesting alternative is \cite{Bodendorfer:2011nv,Bodendorfer:2011nw,Bodendorfer:2011nx}, which shares with this approach an $SO(4)$ phase space. But instead of implementing primary and secondary simplicity constraints, only the primary are implemented. For a consistent dynamics one needs to adjust the Hamiltonian constraints. For a discussion of the various ways and difficulties to implement primary simplicity constraints (which turn out to be non--commuting in the quantum case despite being first class classically) see \cite{Bodendorfer:2011pa}.}



\item[-] Finally, our analysis highlights a basic difference between the $\gamma$-dependence argument in the continuum and this one in the discrete. In the continuum, the $\gamma$-dependent description of the reduced phase space is classically equivalent the $\gamma$-independent one.  They are two parameterizations of the same phase space, related by a canonical transformation, albeit a highly non-trivial one. In the discrete, the $\gamma$-independent Regge geometry phase space is truly a subspace of the $\gamma$-dependent Twisted geometry one.  Furthermore, all the 4d-dihedral angles commute with respect to the Dirac bracket, while their continuum counterpart in the $\gamma$-independent description, the Lorentz connection, is non-commutative. 

\end{itemize}


\section{Plebanski's action and phase space}
\label{sec:plebanski}

The {\sc Plebanski action} for gravity on a smooth manifold $\cspacetime$ takes the form:
\be\label{eq:plebanski-action}
\cS_{Pleb,\cM}[X, w, \phi] = \frac{1}{2}\int_\cspacetime d^4x \left[ \epsilon^{\lambda\mu\nu\rho}\; <{}^{(\gamma)\!} X_{\lambda\mu}, F[w]_{\nu\rho}> + \frac{1}{2}\,\phi^{\lambda\mu\nu\rho} \prec X_{\lambda\mu}, X_{\nu\rho}\succ\right]\;.
\ee
We have introduced co-ordinate charts on space-time parameterized by $x^\mu = (t, x^i)$, where $\mu \in \{0,\,1,\,2,\,3\}$ and $i\in\{1,\,2,\,3\}$. The dynamical fields are  an $\so(4)$-valued bi-vector field $X$ and an $\so(4)$-connection $w$. $<\,,>$ and $\prec\,,\succ$ are the two bilinear forms on $\so(4)$.\footnote{The curvature of this connection is $F[w]_{\mu\nu}{}^A = \partial_{[\mu}w_{\nu]}{}^A + C^{A}{}_{BC}\,w_{[\mu}{}^{B} \,w_{\nu]}{}^C$, where $C^{ABC}$ are the structure constants of $\so(4)$. Meanwhile, the bilinear forms are defined:
\be
<X\,,Y> = X^A \; \delta_{AB}\, Y^{B} \quad\textrm{and}\quad \prec X , Y \succ\, = X^A \, \epsilon_{AB} \, Y^{B} 
\ee}  Moreover, the action contains a map ${}^{(\g)} : \so(4) \rightarrow \so(4) = \delta + \frac{1}{\g}\epsilon$  where $\g\in\R$ is the Barbero-Immirzi parameter.  Thus, the first term in the action describes a topological theory known as $BF$ theory (with Holst modification \cite{Holst:1995pc}). The Lagrange multiplier $\phi^{\lambda\mu\nu\rho}$ is responsible for  imposing the {\sc Lagrangian simplicity constraints}. 
 
As we outlined in the introduction, to perform a phase space path integral quantization of the above system, one needs more information about the constraint structure of the above theory.  In fact, a detailed analysis of the canonical structure has been performed \cite{Buffenoir:2004vx, Alexandrov:2006wt}, including an explicit symplectic reduction.  One finds that at the outset, the constraint structure is rather intricate, but after a reduction with respect to a subset of the constraints, one arrives at a more familiar phase space $\cP_\cspace$ (where $\cspace$ is the characteristic leaf of the space-time foliation: $\cspacetime = \R\times \cspace$).  Its parameterization and symplectic structure are: 
\be
\label{eq:continuum-parameters}
\left(w_i{}^A(\vec x),{}^{(\g)}\Pi^j{}_B(\vec y) \equiv \epsilon^{0ijk}\,\delta_{BC}\,{}^{(\g)\!}X_{jk}{}^C(\vec y)\right) \quad\quad \textrm{and} \quad\quad \left\{w_i{}^A(\vec x),{}^{(\g)}\Pi^j{}_B(\vec y)\right\}  = \delta_i{}^j\, \delta^A{}_B\, \delta^{(3)}(\vec x, \vec y)\;,
\ee
Thus, at this stage of the proceedings, we have the phase space of $\SO(4)$-$BF$ theory.  However, the phase space of gravity is a hypersurface within this phase space.  In other words,
these phase space parameters are still subject to the following constraint system:\footnote{$h_{ij}$ is the spatial metric defined by: $h h^{ij} =\; < \Pi^i,\Pi^j>$, $h = \det (h_{ij})$ where $h^{ij}$ is the inverse of $h_{ij}$.}
{\renewcommand{\arraystretch}{2}\be
\label{eq:continuum-constraints}
\ba{r | rcl}
\hline
\hline

\textsc{Gauss:} &\cG_A &= & \covd_i\, \Pi^{i}{}_A\;,\\
 
 \textsc{Primary simplicity:} & \Phi^{ij} &=& \prec \Pi^i, \Pi^j\succ\;,\\
 
 \textsc{Secondary simplicity:}& \Psi^{ij} &=&  \epsilon^{0mn(i} <h_{mp}\; {}^{(\g)}\Pi^p, D_n\Pi^{j)} >  \;,\\ \hline
 
 \textsc{Vector:} & \lambda_i&=&  \epsilon_{0ijk}\, \Pi^j{}_A\,\delta^{AB}\,F[w]^k{}_B \;,\\

 \textsc{Scalar:} & \lambda_0&=&   h_{ij}\,	\Pi^{i}{}_A\, \epsilon^{AB}\, F[w]^j{}_B\;.\\

 \hline
 \hline
 \ea
\ee}

We have split the constraints here. The vector and scalar constraints are related to the imposition of 4-dimensional diffeomorphism symmetry, while the Gauss, primary and secondary simplicity constraints reduce to a 3-dimensional geometrical phase space. We shall impose the Gauss and simplicity constraints in the discrete setting, while leaving the dynamical constraints for future consideration.  (In fact, in previous work \cite{Dittrich:2008ar}, we have examined the imposition of a more tractable \lq\lq flat dynamics'' constraint.)

\medskip

\noindent{\bf Remark:} For completeness, we note here that this phase space is essentially the same as that of the Holst-modified Einstein-Cartan action \cite{Alexandrov:2006wt}.  Moreover, one arrives at the phase space of LQG in the following way: one gauge-fixes the boost part of the above Lorentz Gauss constraint, which allows one to explicitly solve all the simplicity constraints in terms of an $\su(2)$-connection $A$ and a triad field $E$. The parameters and symplectic structure on this phase space are:
\be
\Big(A_i{}^a(\vec{x}) \equiv \Gamma_i{}^a[E(\vec{x})] - \gamma K_i{}^a(\vec{x}),\, E^j{}_b(\vec{y}) \equiv \Pi^j{}_{0b}(\vec{y}) \Big) \quad \quad \textrm{and} \quad\quad \Big\{A_i{}^a(\vec{x}), E^i{}_b(\vec{y}) \Big\} = \gamma\,\delta_i{}^j\,\delta^a{}_b\,\delta^{(3)} (\vec{x},\vec{y})\,. 
\ee
where $\Gamma$ denotes the spatial Levi-Civita connection constructed from the triad and $K$ encodes the extrinsic curvature tensor.
As one can see, this phase space is essentially identical to that of $\SU(2)$-$BF$ theory, although it is still subject to the vector and scalar constraints.  Moreover, it is the discretization of this phase space that leads to twisted geometries.


\section{Discrete Plebanski phase space: $\cP_{\dspace}$}
\label{sec:poisson}

\begin{figure}[htb]
\begin{center}
\includegraphics[scale = 1.2]{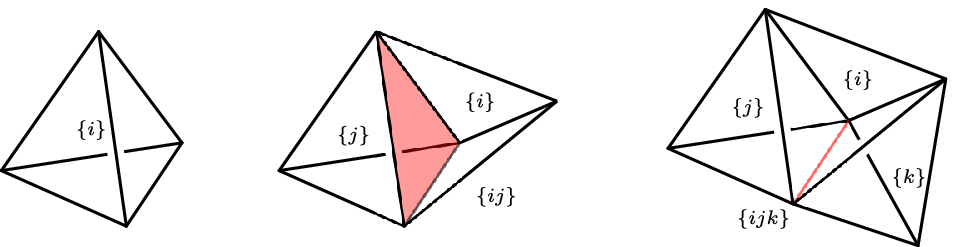}
\caption{\label{fig:indexing} Indexing the various sub-simplices of $\dspace$.}
\end{center}
\end{figure}

As detailed comprehensively in \cite{Dittrich:2008ar, Dittrich:2010ey}, the discretization procedure $\cspace\rightarrow\dspace$ entails the replacement of $\cP_{\cspace}$ with $\cP_{\dspace}$ along with their corresponding constraint sets. On the simplicial complex, tetrahedra are indexed by $\{i\}$. Therefore, triangles (edges) are indexed with respect to the couple (a triple) of tetrahedra to which they belong. We illustrate these choices in Figure \ref{fig:indexing}.

\begin{figure}[hbt]
\centering
\includegraphics[scale = 1.2]{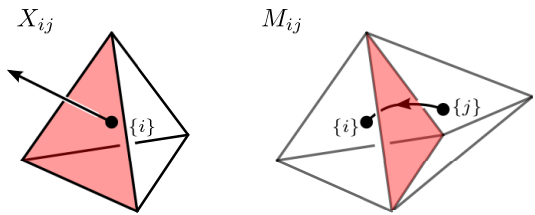}
\caption{\label{fig:parameters-BF} The discrete bi-vector and discrete connection.}
\end{figure}

\noindent  The  variables parameterizing $\cP_{\dspace}$ are a discrete connection and a discrete bi-vector field, denoted by $(M_{ij}, X_{ij})$, respectively. 
We illustrate this in Figure \ref{fig:parameters-BF}.

\noindent The symplectic structure on $\cP_{\Delta}$ is:
\be\label{eq:discrete-sympectic-BF}
\ba{rcl}
\{ M_{ij\pm}{}^{AD}, X_{ij\pm}{}^B\} & = & \dfrac{\gamma}{\gamma \pm 1}\; C^{AB}{}_{C}\; M_{ij\pm}{}^{CD}\;,\\[0.3cm]
\{X_{ij\pm}{}^A, X_{ij\pm}{}^{B}\} & = & \dfrac{\gamma}{\gamma \pm 1}\; C^{AB}{}_{C}\; X_{ij\pm}{}^{C}\;,\\[0.4cm]
\{M_{ij\pm}{}^{AB}, M_{ij\pm}{}^{CD}\} & = & 0\;,
\ea
\ee
where we have separated the degrees of freedom into the self-dual and anti-self-dual sectors using the appropriate projectors on $\so(4)$. More details can be found in Appendix \ref{conv}. 

At this moment, we have an important choice to make concerning the discrete constraint.  To phrase our dilemma differently, what properties should we like the discrete reduced phase space to possess.  Naturally, we would like to arrive ultimately at a discrete geometrical phase space and we take the view that the most appropriate choice is the set of piecewise linear flat geometries - known also as Regge geometries in the literature. We shall take this as our guiding principle when choosing a constraint set.

With this in mind, there are several constraints to implement on this system if one is to remove non-geometrical configurations. They take the form:
\be\label{eq:discrete-constraints}
\ba{rclclcp{5cm}}
\multirow{2}{*}{\phantom{$\Bigg\{$}\textsc{Matching:}} & \multirow{2}{*}{$\Bigg\{$}& {}^{1}\scm_{ij\pm}{}^{A} &:=& X_{ji\pm}{}^A + M_{ji\pm}{}^{A}{}_{B} \; X_{ij\pm}{}^B
&\multirow{2}{*}{$\Bigg\}$}&\multirow{2}{6cm}[-0.15cm]{kill the doubling of variables associated to each triangle,}  \\[0.2cm]

&&  {}^{2}\scm_{ij\pm}{}^{AB} &:=&  (M_{ij\pm}\; M_{ji\pm}){}^{AB} - \delta^{AB} &&  \\[0.2cm]

\textsc{Closure:} &&\scg_{i\pm}{}^A &:=& \sum_{j} X_{ij\pm}{}^A && $\sim \cG^A\;, $\\[0.2cm]

\textsc{Diagonal simplicity} && {}^{}\scd_{ij} & := &  A_{ij+} - A_{ij-} &\multirow{2}{*}{$\Bigg\}$}&\multirow{2}{6cm}[-0.15cm]{$\sim\Phi^{ij}\;,$} \\[0.2cm]

\textsc{Cross simplicity:} && {}^{}\scc_{ijk} & := &  \phi_{ijk+} - \phi_{ijk-} & &\\[0.2cm]

\begin{figure}[htb]
\begin{center}
\includegraphics[scale = 1.2]{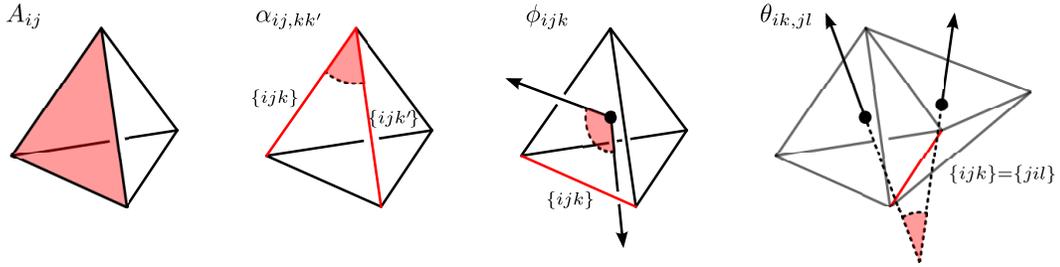}
\caption{\label{fig:variables-gauge} The gauge invariant variables.}
\end{center}
\end{figure}

\textsc{Edge simplicity:} && {}^{}\sce_{ij;kl} &:=&  \theta_{ij;kl+} + \theta_{ij;kl-} & & $\sim\Psi^{ij}\;.$ 
\ea
\ee
The simplicity constraints are written in terms of gauge invariant quantities, detailed in appendix \ref{app:gauge}, which acquire a geometrical meaning on the hypersurface specified by the above simplicity constraints.   $A_{ij}$ denotes the area of the triangle $\{ij\}$, $\cos\phi_{ijk}$ corresponds to the 3d-dihedral angle between the faces $\{ij\}$ and $\{jk\}$, while $\cos \theta_{ij;kl}$ is the 4d-dihedral angle between the tetrahedra $\{i\}$ and $\{j\}$.  We provide an illustration of these quantities for geometrical configurations in Figure \ref{fig:variables-gauge}.  An important point for the following is that there are three edge simplicity constraints and, a priori, three dihedral angles per triangle (and chiral sector). This abundance of different ways to define dihedral angles is indicated in the index structure $\theta_{ij;kl}$, where $ij$ specifies the triangle in question and $kl$ the particular edge which is used for the definition of the dihedral angle. Only on the final reduced (Regge) phase space these different dihedral angles per triangle are forced to agree. 

\noindent{\bf Remark:}  We proved in \cite{Dittrich:2008ar} that this constraint set is sufficient. Namely, upon imposing the simplicity constraints, one arrives at the phase space, in which one can consistently reconstruct edge vectors. Furthermore, these edge vectors are properly parallel-transported by the discrete connection. Thus, in the gauge-invariant subspace, one can reconstruct a unique set of edge-lengths and their conjugate deficit angles; the reduced phase space is the phase space of Regge geometries.

\medskip

\noindent{\bf Remark:} In \cite{Dittrich:2010ey}, we proved that this constraint set, in gauge invariant form (also known as quadratic form of the simplicity constraints), is equivalent to a gauge variant or linear form of the simplicity constraints. In time gauge, this linear form of the simplicity constraints resembles quite obviously a discretization of the primary and secondary simplicity constraints of the continuum, equivalent to the reality conditions for the complex (self-dual) Ashtekar variables, i.e. $E_+=E_-$ and $A_++A_-=\Gamma$. Here, $E$ denotes the triad variables, and $A$ the connection variables in the different duality sectors. $\Gamma$ is the spatial Levi-Civita connection.

\subsection{Discrete constraint analysis}
\label{ssec:discrete-constraints}

One must be careful when referring to the $\theta$-angles as 4d-dihedral angles.   A 4d-dihedral angle, $\theta_{ij}$ is the angle between the normals to the tetrahedra $\{i\}$ and $\{j\}$.  Thus, it is labelled by the shared triangle $\{ij\}$.  Off the constraint hypersurface, however, there are three different ways to define this angle (in terms of the discrete bi-vectors and connection).   Hence, we use the nomenclature $\theta_{ij;kl}$ which denotes the dependence on the edge $\{ijk\} = \{jil\}$. It is only on the geometrical phase space that these three definitions coincide.

Pursuant to this remark, we shall pay some special attention to the simplicity constraints.  It turns out that they have a rather subtle structure.  At the outset, they possess a powerful yet simple appearance imposing:
\begin{center}
 {\sc \lq\lq left-handed geometry $=$ right-handed geometry''}
\end{center}
In fact, they also implicitly ensure that conditions are satisfied among quantities within one and the same chiral sector.  To begin, one can show that there are tidy relationships among the $\theta$-angles associated to a given triangle $\{ij\}$:
\be\label{eq:intrasector}
\theta_{ij;k'l'\pm} - \theta_{ij;kl\pm} = \alpha_{ij;kk'\pm} - \alpha_{ji;l'l\pm}\,.
\ee
The quantities $\theta_{ij;kl}$ and $\theta_{ij;k'l'}$ are two definitions of the 4d-dihedral angle associated to the triangle $\{ij\}$, corresponding to two different edges. In addition, $\alpha_{ij;kk'}$ and $\alpha_{ji,ll'}$ are the two definitions of the same 2d-dihedral angle in the triangle $\{ij\}$, as computed within the two different tetrahedra sharing this triangle. 

Upon imposing the simplicity constraints, one finds that:
\be
\label{eq:gluing}
\left.\ba{rcl}
\phi_{ijk+} &=& \phi_{ijk-} \\[0.2cm]
\theta_{ij;kl+} &=& -\theta_{ij;kl-}\\[0.2cm]
\theta_{ij;k'l'+} &=& -\theta_{ij;k'l'-}
\ea\right\}
\xrightarrow[\textrm{eq. \eqref{eq:intrasector}}]{\textrm{via}}\;
\theta_{ij;k'l'+} = \theta_{ij;kl+} \quad\textrm{and}\quad \alpha_{ij;kk'+} = \alpha_{ji;l'l+}\;,
\ee
which means that on the constraint hypersurface, the $\theta$-angles are really 4d-dihedral angles. They only depend on the triangle $\{ij\}$.  Furthermore, equation \eqref{eq:gluing} shows that if it should suit our purposes,  we may replace some of the edge simplicity constraints with constraints among the $\alpha$-angles in one chiral sector.  They are known as the {\sc gluing constraints} \cite{Dittrich:2008va,Dittrich:2008ar}. 

%





\section{Imposing the constraints}
\label{sec:reduction}

To begin, let us calculate the dimension of the initial phase space. For a generic triangulation $\Delta$, the dimension of $\cP_{\Delta}$ is $24N_f$, where $N_t$, $N_f$ and $N_e$ are the number of tetrahedra, triangles and edges, respectively.  

Our next task is to find an irreducible set of constraints. Given that the reduced phase space is parameterized by Regge geometries, we know that this irreducible set must number: $24N_f - 2N_e$.

{\renewcommand{\arraystretch}{2}
\be
\ba{r | c | c | r l}
\hline\hline
&&\textsc{\# of constraints}& \multicolumn{2}{c}{\textsc{phase space (dimension) after reduction}}\\ \hline

 \textsc{Stage 1:} & {}^{1}\scm_{ij\pm},\;{}^{2}\scm_{ij\pm}  & 12N_f & 12N_f & \\ \hline

 \textsc{Stage 2:} & \scg_{i\pm}  & 6 N_t\; (=3N_f) & 6N_f & \textsc{Gauge-invariant phase space}\\ \hline

& \scd_{ij} & N_f & & \\

 \textsc{Stage 3:} & \scc_{ijk}   & N_f  & 3N_f & \textsc{Twisted geometries}\\
 
& \sce_{ik,jl} & N_f & &\\ \hline

 & \scf_{ijk} & 2N_t\; (=N_f)  & &\\

\textsc{Stage 4:} & \sca & N_f - N_e& 2N_e & \textsc{Regge geometries}\\

& \sct & N_f - N_e & &\\ \hline\hline
 
\ea
\ee
}

As one can see, we have partitioned the constraints into four separate subsets:  {\sc stage 1} involves the matching conditions ${}^{1}\scm_{ij\pm}$ and ${}^{2}\scm_{ij\pm}$;  {\sc stage 2} involves the closure constraint $\scg_{i\pm}$; {\sc stage 3} involves the diagonal, cross and a subset of the edge simplicity constraints;  while {\sc stage 4} involves the rest of the edge simplicity constraints, albeit in an alternative form. We shall detail their exact form later in this section.

\medskip

\noindent{\bf Remark:} As one might imagine, we re-express the final edge simplicity constraints in another form in order to facilitate their imposition.  Note that the gauge-invariant phase space possesses a basis with $6N_f$ elements split half-and-half between the chiral sectors.  The irreducible set of simplicity constraints contains $6N_f - 2N_e$ elements.  Thus, while we can impose $3N_f$ of these in their \lq\lq left-handed = right-handed'' form in {\sc stage 3}, there is a clear obstruction to the imposition of the final $3N_f - 2N_e$ in this manner.  For this reason, we rewrite them as constraints on a single chiral sector, that is, as gluing constraints.

\subsection{{\sc Stage 1:} Matching constraints}

The constraints are 2nd class and couple the $\{ij\}$ variables with $\{ji\}$ variables. As a result, we may deal with them in two equivalent ways:  either we solve the constraints explicitly within the Poisson bracket, or we perform a symplectic reduction to obtain a Dirac bracket.  We choose the former here as the constraints are easily solved: 
\be\label{mat01}
\ba{rclcrcl}
\{M_{ij\pm}{}^{AD}, X_{ij\pm}{}^{B}\} & = & \dfrac{\gamma}{\gamma \pm 1}\; C^{ABC}\; M_{ij\pm}{}^{CD}&\qquad\qquad& \{M_{ij\pm}{}^{DA}, X_{ji\pm}{}^{B}\} & = & \dfrac{\gamma}{\gamma \pm 1}\; C^{ABC}\; M_{ij\pm}{}^{DC},\\[0.3cm]
\{X_{ij\pm}{}^A, X_{ij\pm}{}^{B}\} & = & \dfrac{\gamma}{\gamma \pm 1}\; C^{ABC}\; X_{ij\pm}{}^{C},&&\{X_{ij\pm}{}^A, X_{ji\pm}{}^{B}\} & = & 0,\\[0.4cm]
\{M_{ij\pm}{}^{AB}, M_{ij\pm}{}^{CD}\} & = & 0, &&\{M_{ij\pm}{}^{AB}, M_{ji\pm}{}^{CD}\} & = & 0.
\ea
\ee
The number of matching constraints is $12N_f$ so that the reduced phase space has dimension $12N_f$.

\subsection{{\sc Stage 2:} Closure constraint}

The closure constraints $\scg_{i\pm}^{A}$ are 1st class and generate $\SO(4)$ gauge transformations.  We can take these constraints into account by restricting to the constraint hypersurface and parameterizing the gauge orbits using a basis of gauge-invariant variables.   The number of constraints is $6N_t = 3N_f$ but since they are 1st class, they remove twice as many phase space dimensions. Thus, the resulting gauge-invariant phase space has dimension $6 N_f$. To parameterize it, we shall use the following gauge-invariant parameters:
\be\label{eq:gauge-parameters}
{\renewcommand{\arraystretch}{2}
\ba{rc | c | c}
\hline
\hline
& &  \textsc{\# in $\dspace$} & \textsc{\# used}\\ 
\hline
\textsc{Areas:} & A_{ij\pm} & 2N_f & 2N_f\\
\textsc{3d-dihedral angles:} & \phi_{ijk\pm} & 12N_t & 4N_t = 2N_f\\
\textsc{Averaged 4d-dihedral angles:} & \bar\theta_{ij\pm} & 2N_f &  2N_f  \\
\hline
\hline
\ea}
\ee
In each chiral sector, we use all the areas, two (non-opposite) 3d-dihedral angles from each tetrahedron, and one averaged 4d-dihedral angle per face.  It is a matter of taste that prompts us to use the average of the three 4d-dihedral angles attached to the triangles.  We would rather not single out one of the three as being special.  The rest of the gauge-invariant quantities can be rewritten in terms of these chosen parameters, so we have a basis (see Appendix \ref{app:gauge-basis} for details). 

To keep matters clear, let us present just a schematic of the symplectic structure on the gauge-invariant phase space: 


%


\be\label{eq:gauge-algebra}
{\renewcommand{\arraystretch}{3}
\ba{rcl | rcl | rcl}
\hline
\hline
\{A_{\pm}, A_{\pm}\} &=& 0 &

 \{A_{\pm}, \phi_{\pm}\} &=& 0 &

 \{\phi_{\pm},\phi_{\pm}\} &=& \pm\; 2\dfrac{\gamma}{\gamma\pm 1} \dfrac{\sin\alpha_{\pm}}{A_{\pm}} \\
 
\{A_{\pm}, \bar\theta_{\pm}\}  &=& \pm\; 2\dfrac{\gamma}{\gamma\pm 1}  &

\{ \phi_{\pm}, \bar\theta_{\pm}\}
&=& \pm\; 2 \dfrac{\gamma}{\gamma\pm 1} \dfrac{f_{(1)}(\{\phi_{\pm}\})}{A_{\pm}} &

\{\bar\theta_{\pm}, \bar\theta_{\pm}\} &=& \pm\;2\dfrac{\gamma}{\gamma\pm 1} \dfrac{f_{(2)}(\{\phi_{\pm}\})}{A_{\pm}}\\[0.3cm]
\hline
\hline
\ea}\nonumber
\ee
where we present a much more comprehensive list in Appendix \ref{app:gauge}.

\subsection{{\sc Stage 3:} \lq\lq left-handed = right-handed'' simplicity constraints}
\label{asimp}

The simplicity constraints $\scd_{ij}$, $\scc_{ijk}$ and $\sce_{ik,jl}$ are 2nd class.   As we detailed earlier, we split them into two subsets.  We impose the first subset at this stage.  They enforce that the left-handed basis elements  match up with their right-handed counterparts in the following fashion: 
\be\label{eq:simplicity-one}
{\renewcommand{\arraystretch}{2}
\ba{r | lcl | c | c}
\hline
\hline
& & && \textsc{\# in $\dspace$} & \textsc{\# used}\\ 
\hline
\textsc{Diagonal simplicity:} & \scd_{ij} &=& A_{ij+} - A_{ij-} & N_f & N_f\\
\textsc{Cross simplicity:} & \scc_{ijk} &=& \phi_{ijk+} - \phi_{ijk-} & 6N_t & 2N_t = N_f\\
\textsc{Partial edge simplicity:} & \bar\sce_{ij} &=& \bar\theta_{ij+} + \bar\theta_{ij-} & 3N_f &  N_f  \\
\hline
\hline
\ea}
\ee
Note again that we utilize that subset of the edge simplicity constraint corresponding to the averaged 4d-dihedral angles. This simply halves the dimension of the phase space down to $3N_f$.

We also wish to have the symplectic structure on the reduced phase space, which involves calculating the Dirac bracket: 
\be\label{eq:loop-bracket}
\{\cdot, \cdot\}_L = \{\cdot,\cdot\} - \{\cdot, \cl_I\}\, \dli_{IJ} \,\{\cl_J,\cdot\}\;,
\ee
where $\cl_I$ denotes elements of the above constraint set  and $\dl_{IJ} = \{\cl_I,\cl_J\} $ is the corresponding Dirac matrix.  We present the details in Appendix \ref{app:dirac}, and give a summary here. It emerges that for the basis parameters, the Dirac bracket is merely a rescaling of the corresponding Poisson structure: 
\be\label{eq:loop-algebra}
{\renewcommand{\arraystretch}{3}
\ba{lclcl | lclcl }
\hline
\hline
\{A_{+}, A_{+}\}_L &=& \{A_{+}, A_{+}\} &=&  0 & \{A_{+}, \bar\theta_{+}\}_L  &=& \dfrac{\gamma+1}{2\gamma}\{A_{+}, \bar\theta_{+}\}  &=&  1\\

 \{A_{+}, \phi_{+}\}_L &=&  \{A_{+}, \phi_{+}\} &=& 0 & \{ \phi_{+}, \bar\theta_{+}\}_L &=& \dfrac{\gamma+1}{2\gamma}\{ \phi_{+}, \bar\theta_{+}\} &=&  \dfrac{f_{(1)}(\{\phi_{+}\})}{A_{+}} \\

 \{\phi_{+},\phi_{+}\}_L &=& \dfrac{\gamma+1}{2} \{\phi_{+},\phi_{+}\} &=&  \gamma \dfrac{\sin\alpha_{+}}{A_{+}} & \{\bar\theta_{+}, \bar\theta_{+}\}_L &=& \dfrac{\gamma+1}{2\gamma^2}\{\bar\theta_{+}, \bar\theta_{+}\} &=& \dfrac{1}{\gamma} \dfrac{f_{(2)}(\{\phi_{+}\})}{A_{+}}\\[0.3cm]
\hline
\hline
\ea}
\ee

\medskip

\noindent{\bf Remark:} Let us reiterate that the commutation relations presented above are among the basis parameters and as such, they contain all the information about the symplectic structure on the reduced phase space.  One should not expect, however,  that the Dirac bracket among more general quantities is a simple rescaling of its corresponding Poisson bracket.  This stems from the fact that for the basis parameters the brackets are rescaled in different ways. To be a bit more explicit, let us calculate the Dirac bracket between a 3d-dihedral angle $\phi_+$ and a 4d-dihedral angle $\theta_+$ (not the average, rather one with dependence on an edge of the shared triangle): $\{\phi_+, \theta_+\}_L$.  We know that $\theta_+ = \bar\theta_+ + \sum \alpha_+$ and that every $\alpha_+$ may be written as a function of the 3d-dihedral angles.\footnote{To be a bit more precise, we are considering the situation: $\{\phi_{ijk+}, \theta_{ij,kk'+}\}_L$, where we may write:
 $$\theta_{ij;kl+} = \bar\theta_{ij_+} \underbrace{- \alpha_{ij;kk'} + \alpha_{ij;kk''} + \alpha_{ji;l'l} - \alpha_{ji;l''l}}_{+ f(\{\phi_+\})}\;.$$  } Thus:
\be\label{eq:non-basis}
{\renewcommand{\arraystretch}{2}
\ba{rcl}
\{\phi_+, \theta_+\}_L &=& \{\phi_+, \bar\theta_+\}_L + \{\phi_+, f(\{\phi_+\})\}_L \\
&=& \dfrac{\gamma+1}{2\gamma}\{\phi_+, \bar\theta_+\} + \dfrac{\gamma+1}{2} \{\phi_+, f(\{\phi_+\})\}\\
&=& \dfrac{\gamma+1}{2\gamma}\{\phi_+, \theta_+\} + \dfrac{\gamma^2 - 1}{2\gamma} \{\phi_+, f(\{\phi_+\})\}\;.
\ea}
\ee
So the Dirac brackets involving non-basis 4d-dihedral angles are not simple rescalings of their Poisson bracket counterparts.
Having said that, one finds that analogous commutation relations occur in the loop gravity phase space below.

\subsubsection{Relation to loop gravity phase space}  
 We claim now that we are at the phase space of loop gravity restricted to the graph topologically dual our triangulation $\Delta_3$.  To cut to the chase, the subset of the loop gravity phase space is parameterized by a discrete triad and discrete $\su(2)$ connection: $(x_{ij}{}^r,\, m_{ij}{}^{st})$. With regards to the continuum theory, $m_{ij}$ is the holonomy formed from the Ashtekar-Barbero connection.   
The symplectic structure is then given by:
\be
\ba{rcl}
\{m_{ij}{}^{ru}, x_{ij}{}^s\}_{LG} &=& \g\,\epsilon^{rs}{}_{t}\; m_{ij}{}^{tu},\\[0.2cm]
\{x_{ij}{}^r, x_{ij}{}^{s}\}_{LG} &=& \g\,\epsilon^{rs}{}_{t}\; x_{ij}{}^{t},\\[0.2cm]
\{m_{ij}{}^{ru}, m_{ij}{}^{sv}\}_{LG} &=& 0.
\ea
\ee
As usual, we reduce to the $\su(2)$ gauge-invariant subspace.  This is $3N_f$ dimensional and a basis is parameterized by:
\be\label{eq:lg-gauge-parameters}
{\renewcommand{\arraystretch}{2}
\ba{rc | c | c}
\hline
\hline
& &  \textsc{\# in $\dspace$} & \textsc{\# used}\\ 
\hline
\textsc{Areas:} & A_{ij}  & N_f & N_f\\
\textsc{3d-dihedral angles:} & \phi_{ijk\pm} & 6N_t & 2N_t = N_f\\
\textsc{Averaged 4d-dihedral angles:} & \bar\theta_{ij\pm} & N_f &  N_f  \\
\hline
\hline
\ea}
\ee
These $\su(2)$ quantities are defined in an analogous fashion to their $\so(4)$ counterparts earlier, and we have utilized identical labels to facilitate comparison of the commutation relations \eqref{eq:loop-algebra} and \eqref{eq:lg-loop-algebra}.   Upon calculating the commutation relations among basis parameters, one finds that:
\be\label{eq:lg-loop-algebra}
{\renewcommand{\arraystretch}{3}
\ba{lcl | lcl }
\hline
\hline
\{A_{+}, A_{+}\}_{LG}  &=&  0 & \{A_{+}, \bar\theta_{+}\}_{LG}   &=&   \gamma\\

 \{A_{+}, \phi_{+}\}_{LG} &=& 0 & \{ \phi_{+}, \bar\theta_{+}\}_{LG}  &=&  \gamma\dfrac{f_{(1)}(\{\phi_{+}\})}{A_{+}} \\

 \{\phi_{+},\phi_{+}\}_{LG}  &=&  \gamma \dfrac{\sin\alpha_{+}}{A_{+}} & \{\bar\theta_{+}, \bar\theta_{+}\}_{LG}  &=& {\gamma} \dfrac{f_{(2)}(\{\phi_{+}\})}{A_{+}}\\[0.3cm]
\hline
\hline
\ea}
\ee
As one can see, these relations are quite similar to the ones we found in \eqref{eq:loop-algebra}.  There is a difference in their scaling w.r.t.\ $\gamma$, however, which can be explained in the following fashion. The LQG connection involves $\gamma$, schematically: $A=\Gamma+\gamma K$, whereas we started with an $\SO(4)$ connection that was $\gamma$-independent.  As a result, the dihedral angle $\bar\theta_{LG}$ also involves $\gamma$.  Indeed, replacing $\bar \theta_{LG} \rightarrow \frac1\gamma \bar\theta_{LG}$, one subsequently finds agreement between the brackets \eqref{eq:loop-algebra} and \eqref{eq:lg-loop-algebra}. 

Moreover, in the loop gravity phase space, it is rather the rescaled 4d-dihedral angles $\frac1\gamma \theta_{LG}$ that are related via the 2d-dihedral angles $\alpha_{LG}$. Schematically: $\frac1\gamma\bar\theta_{LG} = \frac1\gamma\theta_{LG} + \sum \alpha_{LG}$.  Thus, one finds that the commutation relations involving non-basis 4d-dihedral angles are identical to those in \eqref{eq:non-basis}.

\subsection{{\sc Stage 4:} gluing constraints}
\label{ssec:gluing}

It is now time to deal with the remaining $3N_f -2N_e$ edge simplicity constraints, which we refer to in their many forms as gluing constraints.   It has been shown \cite{Dittrich:2010ey} that although the edge simplicity constraints relate chiral sectors, they may been rewritten as relations among objects of one and the same sector.  Imposing them is equivalent to imposing:
\be
\label{eq:bsimp01}
\theta_{ij;kl+} = \theta_{ij;k'l'+} \quad\quad\textrm{or}\quad\quad \alpha_{ij;kk'+} = \alpha_{ji;l'l+}\,.
\ee
In words, the left hand relation states that the different definitions for the 4d-dihedral angles agree, while the right hand one states that the definitions for the 2d-dihedral angles agree.
The problem is that all the above variables are complicated functions of the basis parameters and thus rather tricky to impose, even via symplectic reduction.  One may devise yet another equivalent set of constraints by the following argument.  

So far, the 3d-geometrical information has been encoded in both the area variables {\it and} the 3d-dihedral angles $\phi_{ijk}$. However, for discrete 3d-geometries, the area variables provide already an over-complete set. Thus, the 3d-dihedral angles on the geometric constraint hypersurface are actually determined as function of the areas:
\be\label{fcons}
\scf_{ijk} =\phi_{ijk+} - \Phi_{ijk+}[\{A\}] \;.
\ee
The advantage of rewriting the constraints in this manner is that the Poisson (or Dirac) brackets between these constraints is straightforward to determine. Moreover, it leads to a Dirac matrix that factorizes over tetrahedra. In other words, these constraints are second class, while the two constraints per tetrahedron that fix the two independent dihedral angles form conjugate pairs.

For the discrete spatial manifold described by the boundary of a 4-simplex, these are a necessary and sufficient set of constraints to reduce the phase space to the geometrical one.  However, for regular triangulations that are larger than the boundary of the 4-simplex, there are generally more area variables than length variables. Consequently, the constraints \eqref{fcons} are necessary but insufficient for such triangulations and one requires additional constraints that fix the areas as functions of the length variables. These latter constraints can be reconstructed using our basis variables if the full set of simplicity constraints holds \cite{Dittrich:2008ar}: 
\be\label{aco}
\sca : A_{ij+} = a_{ij}[\{l(A)\}] \, .
\ee
 Here $l(A)$ denotes some particular choice of how to construct the lengths from the areas $A_+$, while $a_{ij}(l)$ is the standard area of a triangle $\{ij\}$ expressed as a function of its three edge-length variables.
As we shall comment below, this set of constraints is 1st class, even Abelian, with respect to the Dirac brackets - both the brackets in \eqref{eq:loop-algebra} and the Dirac brackets after imposing the constraints \eqref{fcons}. 

 Thus, we now have an alternative set of $2N_f - N_e$ constraints, of which a subset of cardinality $N_f - N_e$ is 1st class, that capture the remaining $3N_f - 2N_e$ edge simplicity constraints:  
\be\label{eq:gauge-parameters}
{\renewcommand{\arraystretch}{2}
\ba{r | lcl | c | c}
\hline
\hline
& & && \textsc{\# in $\dspace$} & \textsc{\# used}\\ 
\hline
\textsc{Gluing:} & \scf_{ijk} &=&\phi_{ijk+} - \Phi_{ijk+}[\{A\}] & 6N_t & 2N_t = N_f\\
\textsc{Area-length:} & \sca &:&A_{ij+} = a_{ij+}[\{l\}] & N_f - N_e& N_f - N_e \\
\hline
\hline
\ea}
\ee

\subsubsection{Case 1: Boundary of a 4-simplex}

In this scenario, 
we are left with just the $\scf$ constraints on the 3d-dihedral angles.  There are $2N_f = 10$ such constraints, which rewrite the 10 basis 3d-dihedral angles in terms of the areas.  The bracket on the reduced phase space is:
\be\label{eq:regge-bracket}
\{\cdot, \cdot\}_R = \{\cdot,\cdot\}_L - \{\cdot, \scf\}_L\, \dri \,\{\scf,\cdot\}_L\;,\nonumber
\ee
where $\dri$ is the Dirac matrix corresponding to the constraint set $\scf$.  With a little effort, see Appendix \ref{app:stage-four}, one can show that:
\be\label{eq:regge-algebra}
{\renewcommand{\arraystretch}{3}
\ba{lclcl | lclcl }
\hline
\hline
\{A_{+}, A_{+}\}_R &=& \{A_{+}, A_{+}\}_L &=&  0 & \{A_{+}, \bar\theta_{+}\}_R  &=& \{A_{+}, \bar\theta_{+}\}_L  &=&  1\\

 \{A_{+}, \phi_{+}\}_R &=&  \{A_{+}, \phi_{+}\}_L &=& 0 & \{ \phi_{+}, \bar\theta_{+}\}_R &=& \{ \Phi_{+}[\{A\}], \bar\theta_{+}\}_R &=& \dfrac{\partial \Phi_+[\{A\}]}{\partial A} \\

 \{\phi_{+},\phi_{+}\}_R &=& 0 & &    & \{\bar\theta_{+}, \bar\theta_{+}\}_R &=& \multicolumn{3}{l}{ \{\bar\theta_{+}, \bar\theta_{+}\}_L - \{\bar\theta_+,\scf\}_L  \dri \{\scf, \bar\theta_+\}_L  } \\[0.3cm]
\hline
\hline
\ea}\nonumber
\ee
We note that we have made significant progress on all the commutation relations except the final one containing two 4d-dihedral angles.  One can make yet more progress via formal manipulations for this last bracket.  However, ultimately one is forced to explicitly calculate a matrix inverse.  Even for the simple case of a generic 4-simplex boundary, this is rather involved, see Appendix \ref{app:stage-four}. A full analytical evaluation is not possible, as this would require an analytical expression for the length of one simplex expressed as a function of its areas. Such a general expression is not available, as this would involve the general solution of a higher order polynomial.
 Having said that, we performed an explicit evaluation of the matrices in the equilateral case (see Appendices \ref{app:stage-four} for details) and found that it vanishes:
\be
\{\bar\theta_{+}, \bar\theta_{+}\}_R     = 0\;.
\ee
We can devise another argument to show that this bracket must vanish. On the initial $\SO(4)$ phase space, one may impose flatness constraints $F_{e}=\text{id}$, demanding that holonomies around edges are equal to the identity. Such constraints impose the dynamics of $BF$ theory and are 1st class.  Geometrically, the flatness constraints generate translations in the bi-vectors $X$, that is, they change the spatial (geometric) configurations. Then, let us reduce the initial $\SO(4)$ phase space to the one only allowing geometric configurations, that is, to the one in which we can consistently reconstruct the edge-length variables and dihedral angles. For the boundary of the 4-simplex, we know that (4d) flatness can still be imposed and that the constraints take the form \cite{Dittrich:2008pw, Dittrich:2008ar}:
\be\label{fconst}
C_{ij} =\cos \bar \theta_{ij+}  -\cos \Theta_{ij}( \{A_{ij+}\})\;,
\ee
where $ \Theta_{ij}( \{A_{ij+}\})$ is the geometric 4d-dihedral angle as computed from the areas of a flat 4-simplex.\footnote{There are global and discrete ambiguities in reconstructing the geometry of a simplex from the areas. Here however, we need only local reconstructibility.} The flow of the flatness constraints is now restricted to geometric configurations. Indeed, it leads to 10-dimensional gauge orbits. These correspond to the possible changes in the 10 edge-lengths, along with the associated changes in the 4d-dihedral angles, subject to keeping the 4d-geometry of the simplex flat.  In order to produce a consistent flow, the constraints \eqref{fconst} need to be 1st class. This is the case if and only if the 4d-dihedral angles commute:
\be
\{C_{ij},C_{kl}\}_R\, \simeq \, \sin \bar\theta_{ij+} \sin \bar\theta_{kl+} \left( \frac{\partial \Theta_{ij}(A)}{\partial A_{kl+}} -\frac{\partial \Theta_{kl}(A)}{\partial A_{ij+}}   + \{ \bar\theta_{kl+}, \bar\theta_{ij+} \}_R \right) \quad .
\ee
Here the equality sign $\simeq$ indicates an equation which holds on the (flatness) constraint hypersurface. The first two terms cancel each other, due to the Schl\"afli identity, which ensures that $S_\sigma=\sum_{ij} A_{ij+} \Theta_{ij}(A)$ is a generating function for the dihedral angles, that is, $\delta S_\sigma = \sum_{ij} \delta A_{ij+} \Theta_{ij}(A)$.

As a consequence, the existence of a consistent reduction of the flatness constraints to the geometric sector requires that the 4d-dihedral angles commute. This argument can be extended to a class of triangulations, called stacked spheres, who for arbitrary edge length (satisfying appropriate triangle inequalities) are locally embeddable into flat space. Furthermore it also holds for the variables associated to the edges adjacent around a four--valent vertex, as the constraints for such four--valent vertices again agree with constraints imposing flatness \cite{Dittrich:2011ke}.

\subsubsection{Case 2: General triangulation}

For a more general triangulation, we have constraints among the areas, in addition to the constraints fixing the 3d-dihedral angles as functions of the length variables. Their number can be easily determined as the difference between the numbers of triangles and edges in a given 3d triangulation. 

These constraints involve only the area variables, which commute with each other and are therefore 1st class. Thus, we can reduce with respect to these constraints by finding appropriate observables that are invariant under the flow of these constraints. The phase space will then be co-ordinatized by these observables and the induced symplectic structure leads to the same bracket as before $\{\cdot,\cdot\}_R$. One class of observables is obviously given by the edge-length variables $l_e(A)$ associated to the edges $e$ of the triangulation.\footnote{Here, we shall change notation slightly to facilitate our argument: $e$ and $t$ refer to the edges and triangles of the triangulation, respectively.} Nonetheless, one must describe how to compute these edge-length variables in terms of the areas - there will be different possibilities but by definition these will lead to the same result on the area constraint hypersurface. A second class of observables is given by:
\be
p_e=\sum_{t\subset e} \frac{\partial a_t(l)}{\partial l_e} \bar \theta_{t+}\;.
\ee
Here $a_t(l)$ is the area of the triangle $t$ expressed as a function of the edge-lengths, where these edge-lengths are determined by our area variables $A_+$. $\bar \theta_{t+}$ is the dihedral angle associated to the triangle $t$. As can be easily checked, the observables $p_e$ commute with the area constraints (\ref{aco}) and moreover, they are conjugate to the length variables $l_e$ with respect to the brackets $\{\cdot,\cdot\}_R$.  In the case that the dihedral angles to commute with each other, one finds that the momenta $p_e$ also commute:
\be
\{p_e,p_{e'}\}_R \simeq \sum_{t,t'}  \left( \frac{\partial a_t}{\partial l_{e'}\partial l_{e}} \bar \theta_{t+} - \frac{\partial a_{t'}}{\partial l_{e'}\partial l_{e}} \bar \theta_{t'+} \,\,+ \frac{\partial a_t}{\partial l_e} \frac{\partial a_{t'}}{\partial l_{e'}} \{\bar \theta_{t+},\bar\theta_{t'+}\}_R\right)  \; .
\ee
The first two terms cancel each other, as the area derivative factors are only non--zero if the triangles $t,t'$ include both edges $e,e'$, that is $t=t'$. 

Let us note that one finds the same kind of momenta $p_e$ in a canonical framework for Regge calculus \cite{Dittrich:2011ke}, in which the Regge action serves as the generating function for the momenta. There the momenta $p_e$ commute by construction. 

Thus, starting from an $\SO(4)$ phase space and implementing the Gau\ss\ and simplicity constraints,  we found so far a reduced phase space parametrized by length variables and conjugated momenta, which agree in their geometric interpretation with the canonical momenta found for Regge calculus. Also, the commutation relations between length variables and momenta agree with those derived from a Regge phase space. What is more difficult to show directly is that the momenta commute. However, we can also argue here in a somewhat similar fashion to the previous case: since we found that the geometric interpretation of the variables in our reduced phase space and matched those in the Regge phase space, then the dynamics should also be imposed in the same way.  In the general case, the dynamics can be described by discrete time evolution steps that are in fact canonical transformations \cite{Dittrich:2011ke}. For the special situations where topological constraints insist on a flat dynamics, we argued that its consistent implementation required commuting momenta.  Here, we conjecture that there is a unique symplectic structure preserved by the canonical dynamics - in this case, it is the one of the Regge phase space, for which the momenta commute.

\section{Discussion}

Let us briefly recapitulate.  We used a classical phase space approach to shed light on the role played by the Barbero-Immirzi parameter in discrete quantum gravity.  In the continuum description, this parameter appears in a canonical transformation, which turns the momenta conjugate to the triad variables into (commutative) connection variables. However, imposing primary and secondary simplicity constraints in the continuum, one can construct alternative (but non-commutative) connection variables \cite{Alexandrov:2000jw}. It is argued in \cite{Alexandrov:2000jw} that a path integral quantization based on these connections leads to a quantum theory, in which the Barbero-Immirzi parameter does not appear. 

We started from a discrete classical phase space. Introducing a discretization, we replace connection variables by holonomies. This is also done in the (continuum) loop quantum gravity quantization and is the manner in which the Barbero-Immirzi parameter manifests itself in the quantum theory. 

Indeed, after we impose a subset of the simplicity constraints, we find a phase space that can be mapped to the phase space of twisted geometries, which itself arises if we restrict Loop Quantum Gravity to a fixed graph. The Dirac brackets between basic geometric quantities are $\gamma$-dependent and this leads to the appearance of $\gamma$ in the spectra of geometric operators. This latter fact shows that the quantum theories based on different values of the $\gamma$ parameter are not unitarily equivalent - even if we start from a finite dimensional phase space, as arises here from the restriction to a fixed graph. 

This first set of simplicity constraints imposes that for a given basis of gauge invariant phase space variables, the left handed-basis variables agree with the right-handed basis variables (for the 4d-dihedral angles up to sign). However, this does {\it not} mean that all geometrical variables constructible in the left-handed sector agree with their counterparts in the right-handed sector. It is rather after reducing with respect to a larger set of simplicity constraints that the phase space truly possesses the property: \lq\lq left-handed geometry $=$ right-handed geometry\rq\rq. Moreover, after this further reduction the phase space dimension and Dirac brackets agree with their counterparts in the Regge geometries phase space and do not depend on $\gamma$.

Thus, one has the choice to base quantization on either of these phase spaces. Due to the availability of a connection (or rather group holonomies), the loop quantum gravity-like phase space is given by a certain tensor product of $T^*(SU(2))$ spaces and quantization is straightforward \cite{Freidel:2009nu}. Moreover, the connection to the continuum theory via a projective limit is well developed \cite{Thiemann:2007zz}.  In contrast, although a heuristic quantization of the Regge phase space is certainly possible, the imposition of triangle inequalities on this phase space is an open problem. On the other hand, a discrete (classical) canonical dynamics for the Regge phase space is available \cite{Dittrich:2011ke}, whereas a dynamics for (discrete) classical twisted geometries has so far not been discussed. Indeed, the pertinent question here concerns what kind of dynamics one should impose on the non-geometric configurations, or in other words, whether these non-geometric configurations can be suppressed dynamically.

Since the phase spaces have different sizes, one would expect different quantum theories, even at the dynamical level. In particular, any appearance of $\gamma$ in the final dynamical predictions would be an effect due to the larger phase space of discrete loop quantum gravity, as compared to the Regge phase space.

\newpage


\appendix


\section{Conventions}
\label{conv}

The discretization of 4-dimensional $\SO(4)$ $BF$ theory relies heavily on the isomorphism between the space of bi-vectors (anti-symmetric second order tensors) and the algebra $\so(4)$.  In our conventions, the commutation relations among the generators of the algebra take the form: 
\be\label{conv01}
[J_A,J_B] = C_{AB}{}^{C}\; J_C.
\ee
The indices are $A=a\bar{a}$ where $a,\bar{a} \in \{0,1,2,3\}$.  The structure constants are:
\be\label{conv02}
\ba{rcl}
C^{ABC} &:=& \epsilon^{Ar}{}_{s} \; \epsilon^{Bs}{}_{t}\;\delta^{Ct}{}_{r}\\[0.2cm]
&=&  \delta^{ab}\;\delta^{\bar{a}\bar{b}C} + \delta^{\bar{a}\bar{b}}\;\delta^{abC} - \delta^{a\bar{b}}\;\delta^{\bar{a}bC} - \delta^{\bar{a}b}\;\delta^{a\bar{b}C},
\ea
\ee
where the Kronecker delta and the summation convention on this space are given by:
\be\label{conv03}
\ba{rcl}
\delta^{AB} & := & \delta^{ab}\delta^{\bar{a}\bar{b}} - \delta^{a\bar{b}}\delta^{\bar{a}b},\\[0.2cm]
T^{AB}\;U_{B}{}^{C} & := & \dfrac{1}{2}\; T^{Ab\bar{b}} \;U_{b\bar{b}}{}^{C} \quad\quad \textrm{so that} \quad\quad \epsilon^{AB}\;\epsilon_{B}{}^{C} = \delta^{AC}.
\ea
\ee
The Killing form is: 
\be\label{conv04}
C^{A}{}_{CD} \; C^{BCD} = \delta^{AB}
\ee
 while the Hodge map $\epsilon : \so(4) \rightarrow \so(4)$ so that:
\be\label{conv05}
\epsilon(X)^A = \epsilon^A{}_{B}\; X^B
\ee
Using the Killing form and the Hodge map we can define two independent bilinear forms on $\so(4)$:
\be\label{conv06}
<X,Y>\;:= X^{A}\; \delta_{AB}\; Y^{B}  \quad\quad \mathrm{and} \quad\quad \prec X,Y\succ\;:= X^{A}\; \epsilon_{AB}\; Y^{B}.
\ee
We also define  \lq\lq scalar'' and \lq\lq bi-vector'' products:
\be
X\cdot Y := X^A\, \delta_{AB}\, Y^{B}\, \quad\quad\mathrm{and} \quad\quad (X\times Y)^A := C^{A}{}_{BD} \,\epsilon^{D}{}_C \, X^B\, Y^C.
\ee

\vspace{1cm}

\noindent We can use the Hodge operator to split the algebra into self-dual and anti-self-dual sub-algebrae:
\be\label{conv07}
J_{\pm}{}^A : = P_{\pm}{}^{AB} \; J_B, \quad\quad\textrm{where} \quad\quad P_{\pm}{}^{AB} := \frac{1}{2}(\delta^{AB} \pm\epsilon^{AB}).
\ee
It is straightforward to check that these projectors are orthonormal, that is $P_s^{AB}P_{s'}^{BC}=\delta_{ss'}P_s^{AC}$, furthermore $P_+{}^{AB}+P_-{}^{AB}=\delta^{AB}$. 
We may then proceed to generate all manner of projected quantities:
\be\label{conv08}
\ba{rclcrclcl}
X_\pm{}^A &=& P_\pm{}^{A}{}_{A'}\; X^{A'}&& M_\pm{}^{AB} &=& P_\pm{}^{A}{}_{A'}\;M^{A'B} &= & P_\pm{}^{B}{}_{B'}\;M^{AB'},\\[0.2cm]
C_\pm{}^{ABC} &=& P_{\pm}{}^{A}{}_{A'}\;C^{A'BC} &=& P_{\pm}{}^{B}{}_{B'}\;C^{AB'C} &=& P_{\pm}{}^{C}{}_{C'}\;C^{ABC'} &=& P_{\pm}{}^{A}{}_{A'}\;P_{\pm}{}^{B}{}_{B'}\;P_{\pm}{}^{C}{}_{C'}\;C^{A'B'C'}\;.
\ea
\ee
The following useful identity for the $\so(4)$ structure constants:
\be\label{conv09}
C_\pm{}^{ABC}\; C_\pm{}^{A'B'}{}_{C}=2^3 (P_\pm{}^{AA'}\; P_\pm{}^{BB'} - P_\pm{}^{AB'}\; P_\pm{}^{A'B} ).
\ee
 is in close analogy to the relation $\epsilon^{abc}\epsilon^{a'b'c}=(\delta^{aa'}\delta^{bb'}-\delta^{ab'}\delta^{a'b})$ for the $\SO(3)$ structure constants.

In our conventions, $\epsilon^{\lambda\mu\nu\rho}$ is a totally antisymmetric tensor density of weight 1 with $\epsilon^{0123} = 1$.  On the other hand, in absence of a metric, we define $\epsilon_{\lambda\mu\nu\rho}$ as totally antisymmetric tensor density of weight $-1$ with $ \epsilon_{0123} = 1$.  This means that $\epsilon^{\lambda\mu\nu\rho}\;\epsilon_{\lambda\mu\nu\rho} = 4!$.  Furthermore, we define $\epsilon^{ijk} := \epsilon^{0ijk}$ and $\epsilon_{ijk} := \epsilon_{0ijk}$.



\newpage

\section{Gauge-invariant quantities}
\label{app:gauge}

\subsection{Definitions}
\label{app:gauge-defs}

Let us first detail the gauge-invariant variables that occur in the main text and appendices:
\be\label{}
{\renewcommand{\arraystretch}{2.7}
\ba{r | lcl | lcl }
\hline\hline

\textsc{Areas:}  &A_{ij\pm} &:= & \dfrac{1}{\sqrt{2}} |X_{ij\pm}|\\

\textsc{Squared volumes:} & V_{ijkl\pm} &:=& \big[X_{ij}\cdot(X_{ik}\times X_{il})\big]_{\pm}\\

\textsc{Edge Lengths:} &l_{ijk\pm} &:=& \left[\dfrac{|N_{ijk}|}{\sqrt{2|V_{ijkl}|}}\right]_{\pm}\\

\textsc{2d-dihedral angles:} & 
  \cos\alpha_{ij;kl\pm} & := & \dsty \left[\frac{N_{ijk}\cdot N_{ijl}}{|N_{ijk}|\, |N_{ijl}|}\right]_{\pm} & 
  \sin\alpha_{ij;kl\pm} &:=& \dsty \left[\frac{4\, A_{ij}\, V_{ijkl}}{|N_{ijk}|\,|N_{ijl}|} \right]_{\pm}\\

\textsc{3d-dihedral angles:} &
  \cos \phi_{ijk\pm} &:= &\dsty\left[\frac{X_{ij}\cdot X_{ik}}{2\,A_{ij} \, A_{ik}  }\right]_{\pm} &
  \sin \phi_{ijk\pm} &:= & \dsty \left[ \frac{|N_{ijk}|}{2^{\frac{5}{2}}\, A_{ij}\,A_{ik}}\right]_{\pm}

\\

\textsc{4d-dihedral-angles:} & 
  \cos\theta_{ij;kl\pm} &:= &\dsty\left[\frac{N_{ijk}\cdot M_{ij} N_{jil}}{|N_{ijk}|\, |N_{jil}|}\right]_{\pm} & 
  \sin\theta_{ij;kl\pm} &:=&   \dsty \left[\frac{ 4\, A_{ij}\,(X_{ik}\cdot M_{ij}N_{jil}) } {|N_{ijk}|\,|N_{jil}|}\right]_{\pm}\\[0.5cm]
\hline
\hline


\ea}\nonumber
\ee
where $N_{ijk\pm}:= (X_{ij}\times X_{ik})_{\pm}$.  

\medskip

\noindent{\bf Remark:} We define both the cosine and sine of each angle above.  Since they both involve square roots, we do so in order to specify their signs.    A priori, this is an ambiguity, although it is removed by picking one particular sign consistently. Moreover, for $\sin \theta_{\pm}$, we wish to reflect the fact that $\theta_+ = - \theta_-$ for geometric configurations. Thus, we pick the factor $X_{ik\pm}\cdot M_{ij\pm}N_{jil\pm}$ in the numerator, which satisfies $X_{ik+}\cdot M_{ij+}N_{jil+} =  -X_{ik-}\cdot M_{ij-}N_{jil-}$ on the constraint hypersurface. Also, $\sin \alpha$ is essentially a tri-linear function on phase space and the presence of $V$ (rather than $|V|$) represents the change of sign of $\sin\alpha_{ij;kl}$ as one permutes the indices $k$ and $l$.

\medskip

\noindent{\bf Remark:} Let us also details the evaluation of these quantities for geometric configurations. In this situation, there is a reference frames attached to each tetrahedron. Pick a vertex of the tetrahedron as the origin and use the three edge-vectors $e_{1,2,3}$ emanating from it as the basis.  Taking the discrete bi-vectors to be:
\be
{\renewcommand{\arraystretch}{1}
\ba{rclcrcl}
X_{ij}{}^A &=&  \epsilon^{Ab\bar b}\; e_{1}{}^b\, e_2{}^{\bar{b}}\;, &&
X_{ik}{}^A &=&  \epsilon^{Ab\bar b}\; e_{2}{}^b\, e_3{}^{\bar{b}}\;,\\ [0.5cm]
X_{il}{}^A &=&  \epsilon^{Ab\bar b}\; e_{3}{}^b\, e_1{}^{\bar{b}}\;, & &
X_{im}{}^A &=& - X_{ij}{}^A  - X_{ik}{}^A  - X_{il}{}^A \;,
\ea}
\ee
leads to the following:
\be
\ba{rclcrcl}
A_{ij+} &=& \dfrac{1}{2} |e_1| |e_2| \sin\alpha_{ij,kl+}\;,&& \sin\alpha_{ij,kl+} &=& \sqrt{1 - (e_1\cdot e_2)^2}\;,\\
V_{ijkl+} &=&  18 |v_i|^2 && v_{i}{}^{a}\;, &=& \dfrac{1}{6} \epsilon^{abcd}\, e_1{}^b\, e_2{}^c\, e_3{}^d\;,\\  

\ea
\ee
Note that $v_i{}^a$ satisfies: $|v_i| = \textrm{volume of tetrahedron}$.

\subsection{Providing a basis}
\label{app:gauge-basis}

Given a triangulation $\dspace$, we claim that we can parameterize the gauge-invariant phase space using, in each chiral sector,  all the areas, two 3d-dihedral angles per tetrahedron and  one 4d-dihedral angle per triangle. These is a set with $6N_f$ elements, which is certainly the correct number but let us now show that we can reconstruct the rest of the gauge-invariant quantities from these basic set.

Let us first deal with intrinsic quantities, that is, quantities pertaining to a single tetrahedron marked $\{i\}$, with four faces $\{ij\}$, $\{ik\}$, $\{il\}$ and $\{im\}$.   From our proposed basis, we have the following parameters: 
\be\label{asimp09}
A_{ij},\; A_{ik},\; A_{il},\; A_{im},\; \cos\phi_{ijk},\;  \cos\phi_{ijl}.
\ee 
We suppress the $\pm$ here for ease of reading. Then, the other 3d-dihedral angles are given by:
\be\label{asimp10}
{\renewcommand{\arraystretch}{3}
\ba{rcl}
\cos\phi_{ijm} &=& -\dfrac{A_{ij}}{A_{im}} - \cos\phi_{ijk} \, \dfrac{A_{ik}}{A_{im}} - \cos\phi_{ijl} \, \dfrac{A_{il}}{A_{im}}\;,\\

\cos\phi_{ikm} &=& \dfrac{1}{2}\left(\dfrac{A_{ij}^2}{A_{ik}A_{im}} - \dfrac{A_{ik}}{A_{im}} + \dfrac{A_{il}^2}{A_{ik}A_{im}} - \dfrac{A_{im}}{A_{ik}} + 2\cos\phi_{ijl}\,\dfrac{A_{ij}A_{il}}{A_{ik}A_{im}}\right)\;,\\

\cos\phi_{ilm} &=& \dfrac{1}{2}\left(\dfrac{A_{ij}^2}{A_{il}A_{im}}  + \dfrac{A_{ik}^2}{A_{il}A_{im}} - \dfrac{A_{il}}{A_{im}}- \dfrac{A_{im}}{A_{il}} + 2\cos\phi_{ijk}\,\dfrac{A_{ij}A_{ik}}{A_{il}A_{im}}\right)\;,\\

\cos\phi_{ikl} &=& \dfrac{1}{2}\left(-\dfrac{A_{ij}^2}{A_{ik}A_{il}}  - \dfrac{A_{ik}}{A_{il}}- \dfrac{A_{il}}{A_{ik}} + \dfrac{A_{im}^2}{A_{ik}A_{il}} - 2\cos\phi_{ijk}\,\dfrac{A_{ij}}{A_{il}} - 2\cos\phi_{ijl}\,\dfrac{A_{ij}}{A_{ik}}\right)\;.
\ea}
\ee
Moreover, the 2d-dihedral angles may be constructed from these 3d-dihedral angles:
\be
\cos\alpha_{ij;kl} = \frac{\cos\phi_{ikl} - \cos\phi_{ijk}\,\cos\phi_{ijl} }{\sin\phi_{ijk}\sin\phi_{ijl}}\;.
\ee

We are left now with extrinsic quantities, that is, the 4d-dihedral angles.  Consider a triangle $\{ij\}$. It is a face of two tetrahedra $\{i\}$ and $\{j\}$.  They have faces $\{ij\}, \{ik\}$, $\{il\}$, $\{im\}$ and $\{ji\}$, $\{jk'\}$, $\{jl'\}$, $\{jm'\}$, respectively. Thus, for the triangle $\{ij\}$, there are three 4d-dihedral angles: $\theta_{ij;kk'}$, $\theta_{ij;ll'}$ and $\theta_{ij;mm'}$. Using the relations:
\be
\ba{lcl}
\theta_{ij;ll'} &=& \theta_{ij;kk'} + \alpha_{ij;kl} + \alpha_{ji;k'l'}\;,\\[0.3cm]
\theta_{ij;mm'} &=& \theta_{ij;kk'} + \alpha_{ij;km} + \alpha_{ji;k'm'}\;.
\ea
\ee

\subsection{Commutation relations}
\label{app:gauge-relations}

We provided a very cursory digest of the commutation relations among the gauge-invariant parameters of interest in the main text.  We provide a more complete list here:
\be\label{}
{\renewcommand{\arraystretch}{3}
\ba{rcl}
\hline
\hline

\{A_{ij\pm}, A_{mn\pm}\} &=& 0\;,\\ 

\{A_{ij\pm}, \phi_{mnp\pm}\} &=& 0\;,\\


\{\phi_{ijk\pm},\phi_{ijm\pm}\} &=& 
	\pm\; 2\dfrac{\gamma}{\gamma\pm 1} 
		\dfrac{1}{A_{ij\pm}} \sin\alpha_{ij;km\pm}\;, \\  [0.4cm]

\hline

 \{A_{ij\pm}, \theta_{ij;mn\pm}\} &=& \pm\; 2\dfrac{\gamma}{\gamma\pm 1} \;, \\

\{\phi_{ijk\pm},\theta_{ij;kn\pm}\}&=&
	\pm \; 2 \dfrac{\gamma}{\gamma\pm 1}  
	\Bigg(\dfrac{\csc\phi_{ijk\pm}}{A_{ik\pm}} + \dfrac{\cot\phi_{ijk\pm}}{A_{ij\pm}}\Bigg)\;,\\

\{\phi_{ijk\pm},\theta_{ij;mn\pm}\}&=& 
	\pm\; 2  \dfrac{\gamma}{\gamma\pm 1} 
	\dfrac{1}{A_{ij\pm}} \cot\phi_{ijm\pm} \cos\alpha_{ij;km\pm}\;, \\

\{\phi_{ijk\pm},\theta_{im;jn\pm}\} &=& 
	\mp\; 2  \dfrac{\gamma}{\gamma\pm 1} \dfrac{1}{A_{ij\pm}}\csc\phi_{imj\pm} \cos\alpha_{ij;km\pm}  \;,\\[0.4cm]

\hline

\{\theta_{ij;kl\pm},\theta_{ij;mn\pm}\} &=& \pm\;2\dfrac{\gamma}{\gamma\pm 1} \dfrac{1}{A_{ij\pm}}   \Big(\cot\phi_{ijk\pm} \cot\phi_{ijm\pm} \sin\alpha_{ij;km\pm} + \cot\phi_{jil\pm} \cot\phi_{jin\pm} \sin\alpha_{ji;ln\pm} \Big)\\

\{\theta_{ij;kl\pm}, \theta_{im;jn\pm}\} &=& 
	\mp\;2\dfrac{\gamma}{\gamma\pm 1} 
		\dfrac{1}{A_{ij\pm}} \cot\phi_{ijk\pm}\csc\phi_{imj\pm} \sin\alpha_{ij;mk\pm} \;, \\

\{\theta_{ij;kl\pm}, \theta_{im;kn\pm}\} &=& 
	\mp\;2\dfrac{\gamma}{\gamma\pm 1} 
		\dfrac{1}{A_{ik\pm}} \csc\phi_{ijk\pm}\csc\phi_{imk\pm} \sin\alpha_{ik;jm\pm} \;.\\

\{\theta_{ij;kl\pm}, \theta_{ik;jn\pm}\} &=& 	0   \\[0.4cm]

\hline
\hline
\ea}\nonumber
\ee



\newpage

\section{The calculation of $\{\cdot,\cdot\}_L$ in {\sc stage 3}}
\label{app:dirac}

As we mentioned in the main text, we intend to explicitly calculate, via symplectic reduction,  the Dirac bracket on the phase space reduced by the constraint set $\cl_i = \{\scd, \scc, \bar\sce\}$.   At first glance, this might appear to be a rather arduous task.  The exists a neat fact, however, that simplifies the calculations immensely.  Since the constraints are simple linear equalities between pairs of phase space parameters, there is a remarkable similarity between the commutator of constraints and commutators of phase space parameters, as illustrated in \eqref{eq:bracket-relations}:
\be\label{eq:bracket-relations}
{\renewcommand{\arraystretch}{3}
\ba{rcrcrcr}
 \cA & = & \{\scd,\bar\sce\}_+ & = &\dfrac{2\gamma}{\gamma - 1}\{A_+, \bar\theta_+\}\;,\\
\cB & = & \{\scc,\scc\}_+ & = & - \dfrac{2}{\gamma - 1}\{\phi_+, \phi_+\}\;,\\ 
\cC & = & \{\scc,\bar\sce\}_+ & =& \dfrac{2\gamma}{\gamma - 1}\{\phi_+, \bar\theta_+\}\;,\\
\cD & = & \{\bar\sce,\bar\sce\}_+ & =& -\dfrac{2}{\gamma - 1}\{\bar\theta_+, \bar\theta_+\}\;,
\ea}
\ee
where $\{\cdot,\cdot\}_+$ denotes the Poisson bracket evaluated on the reduced phase space. As a result, one may derive the symplectic structure on the reduced phase space from the block form of the inverse of the Dirac matrix.  The Dirac bracket is defined in \eqref{eq:loop-bracket} as:
\be
\{\cdot, \cdot\}_L = \{\cdot,\cdot\} - \{\cdot, \cl_i\}\, \dli_{ij} \,\{\cl_j,\cdot\}\;,\nonumber
\ee
where:
\be\label{eq:loop-matrix}
\dl = \left(\ba{ccc}
0 &0 & \cA\\[0.2cm]
0 &\cB &\cC\\[0.2cm]
-\cA & -\cC^T& \cD
\ea\right)\;, 
\quad\quad 
\dli =  \left(\ba{ccc}
\cA^{-1}\cE^T\cB^{-1}\cE\cA^{-1} + \cA^{-1}\cG\cA^{-1} & - \cA^{-1}\cE^T\cB^{-1} & -\cA^{-1}\\[0.2cm]
-\cB^{-1}\cE\cA^{-1} &\cB^{-1} &0\\[0.2cm]
\cA^{-1} & 0& 0
\ea\right)
\ee
As a simple example, let us calculate:
\be
{\renewcommand{\arraystretch}{3}
\ba{rcl}
\{A_{\cdot\cdot+},\bar\theta_{\cdot\cdot+}\}_L &=& \{A_{\cdot\cdot+}, \bar\theta_{\cdot\cdot+}\} 
				- \{A_{\cdot\cdot+}, \bar\sce\}(\dli)_{31}\{\scd,\bar\theta_{\cdot\cdot+}\}\\

&=& \dfrac{\gamma-1}{2\gamma}\cA 
- \left(\dfrac{\gamma-1}{2\gamma}\right)^2\cA\;\cA^{-1}\cA \\

&=&\dfrac{\gamma^2 - 1}{4\gamma^2}\cA\\

&=& \dfrac{\gamma + 1}{2\gamma} \{A_{\cdot\cdot+},\bar\theta_{\cdot\cdot+}\}  
\ea}
\ee
The other brackets follow in a similar fashion with the results summarized in \eqref{eq:loop-algebra}.

\newpage

\section{The calculation of $\{\cdot,\cdot\}_R$ in {\sc stage 4}}
\label{app:stage-four}

We shall reduce with respect to the constraints $\{\scf, \sca\}$ for the case of the boundary of a 4-simplex.

\subsection{4-simplex boundary}

For the case of a 4-simplex boundary, the remaining constraints simplify to the 2nd class set $\{\scf\}$, and so the Dirac matrix simplifies to $\dr = \{\phi_+, \phi_+\}$. Thus, its inverse is easily calculable as the matrix elements satisfy:
\be\label{eq:regge-inverse}
\dri_{IJ} = -\dfrac{1}{\dr_{\!\!\!JI}} 
\ee
When explicitly calculating large Dirac matrices, it becomes rather important to choose the constraint set wisely, as this will cut down the amount of work considerably.  An independent subset of constraints was found in \cite{dr2}, which we shall utilize here.   First of all, let us choose a parameterization of our 30-dimensional gauge-invariant phase space:\footnote{We have chosen here the $\theta_ij,kl$ rather than the averaged angles $\bar\theta_{ij}$ for purely historical reasons. }
\be\label{dirac01}
{\renewcommand{\arraystretch}{2}
\ba{llllllllll}
A_{12+}, & A_{13+}, & A_{23+}, & A_{24+}, & A_{34+}, & A_{35+}, & A_{45+}, & A_{41+}, & A_{51+}, & A_{52+},\\

\phi_{125+}, & \phi_{135+}, & \phi_{231+}, & \phi_{241+}, & \phi_{342+}, & \phi_{352+}, & \phi_{453+}, & \phi_{413+}, & \phi_{514+}, & \phi_{524+}, \\

\theta_{12;55+}, & \theta_{13;55+}, & \theta_{23;11+}, & \theta_{24;11+}, & \theta_{34;22+}, & \theta_{35;22+}, & \theta_{45;33+}, & \theta_{41;33+}, & \theta_{51;44+}, & \theta_{52;44+},

\ea}
\ee
Then, an appropriate constraint set is:
\be\label{dirac02}
\ba{llllllllll}

\scf_{125}, & \scf_{135}, & \scf_{231}, & \scf_{241}, & \scf_{342}, & \scf_{352}, & \scf_{453}, & \scf_{413}, & \scf_{514}, & \scf_{524}, \\[0.2cm]

\ea
\ee
Due to the nature of the relation \eqref{eq:regge-inverse}, the inverse of the Dirac matrix is easy to compute:  
\be
\dri = 
{\scriptsize\left(
\ba{cccccccccc}
0 & f_{1} & 0 & 0 & 0 & 0 & 0 & 0 & 0 & 0\\[0.2cm]
-f_{1}& 0 & 0 & 0 & 0 & 0 & 0 & 0 & 0 & 0\\[0.2cm]
0& 0& 0 & f_{2} & 0 & 0 & 0 & 0 & 0 & 0\\[0.2cm]
0& 0& -f_{2}& 0 & 0 & 0 & 0 & 0 & 0 & 0\\[0.2cm]
0& 0& 0& 0& 0 & f_{3} & 0 & 0 & 0 & 0\\[0.2cm]
0& 0& 0& 0& -f_3& 0 & 0 & 0 & 0 & 0\\[0.2cm]
0&0&0&0&0&0& 0 & f_{4} & 0 & 0\\[0.2cm]
0&0&0&0&0&0&-f_4& 0 & 0 & 0\\[0.2cm]
0&0&0&0&0&0&0&0& 0 & f_{5}\\[0.2cm]
0&0&0&0&0&0&0&0&-f_5& 0
\ea
\right),}
\quad\quad
\textrm{where:}\quad\quad
f_{1} := -\frac{A_{15+}}{\gamma \sin\alpha_{15,23+}}.
\ee 
Moreover, in $f_{1+i}$, one shifts all the subscripts by $i\; (\textrm{mod}\; 5)$ in the previous equation.   Then, we have:
\be\label{dirac07}
{\scriptsize
\{\phi_+,\theta_+\}_L = \left(
\ba{cccccccccc}
e_{125} & \ua e_{1253} & 0 & 0 & 0 & 0 & 0 & 0& \ub e_{1254} & 0\\[0.2cm]
\ua e_{1352}& e_{135} & 0 & 0 & 0 & 0 & 0 &\ua e_{1534} & \ub e_{1354} & 0\\[0.2cm]

\ub e_{2315} & 0&e_{231} & \ua e_{2314} & 0 & 0 & 0 & 0 & 0 & 0\\[0.2cm]
 \ub e_{2415} & 0& \ua e_{2413}& e_{241} & 0 & 0 & 0 & 0 & 0 &\ua e_{2145}\\[0.2cm]

0 & 0& \ub e_{3421} & 0&e_{342} & \ua e_{3425} & 0 & 0 & 0 & 0  \\[0.2cm]
0 &\ua e_{3251} & \ub e_{3521} & 0&\ua e_{3524}& e_{352} & 0 & 0 & 0 & 0  \\[0.2cm]

0 & 0 & 0 & 0& \ub e_{4532} & 0&e_{453} & \ua e_{4531} & 0 & 0  \\[0.2cm]
0 & 0 & 0 &\ua e_{4312} &\ub  e_{4132} & 0&\ua e_{4135}& e_{413} & 0 & 0  \\[0.2cm]

0 & 0 & 0 & 0 & 0 & 0& \ub e_{5143} & 0& e_{514} & \ua e_{5142}  \\[0.2cm]
 0 & 0 & 0 & 0 & 0 & \ua e_{5423} & \ub e_{5243} & 0& \ua e_{5241}& e_{524} 
\ea
\right),}
\ee
where:
\be\label{dirac08}
\ba{rcl}
e_{ijk} &=&   \Bigg(\dfrac{\csc\phi_{ijk_+}}{A_{ik+}} + \dfrac{\cot\phi_{ijk+}}{A_{ij+}}\Bigg),\\[0.5cm]

\ua e_{ijkl} &=& -\dfrac{1}{A_{ij+}}  \csc\phi_{ilj+} \cos\alpha_{ijkl+} ,\\[0.5cm]

\ub e_{ijkl} &=&   \dfrac{1}{A_{ij+}}  \cot\phi_{ijl+} \cos\alpha_{ijkl+}.
\ea
\ee
The final bracket to calculate is:
\be\label{dirac10}
\{\theta_+,\theta_+\}_L =
{\scriptsize
\addtolength{\arraycolsep}{-1.1mm}
\left(
\ba{cccccccccc}
0& \ua g_{1235} &  \ub g_{2315} &  \ub g_{2415} & 0 & 0 & 0 & 0& - \ub g_{1254} & - \ub g_{2154}\\[0.2cm]
-\ua  g_{1235} & 0 &  \ub g_{3215} & 0 & 0 & - \ub g_{3152} & 0 & \ub g_{1435} & - \ub g_{1354}& 0\\[0.2cm]

- \ub g_{2315} & - \ub g_{3215}&0& \ua  g_{2341} &  \ub g_{3421} &  \ub g_{3521} & 0 & 0 & 0 & 0 \\[0.2cm]
- \ub g_{2415}& 0&-\ua g_{2341} & 0 &  \ub g_{4321} & 0 & 0 & - \ub g_{4213} & 0 & \ub g_{2541}  \\[0.2cm]

0 & 0& - \ub g_{3421} & - \ub g_{4321}&0& \ua g_{3452} &  \ub g_{4532} &  \ub g_{4132} & 0 & 0  \\[0.2cm]
 0 & \ub g_{3152} & - \ub g_{3521}& 0&-\ua g_{3452} & 0 &  \ub g_{5432} & 0 & 0 & - \ub g_{5324} \\[0.2cm]

0 & 0 & 0 & 0& - \ub g_{4532} & - \ub g_{5432}&0& \ua g_{4513} &  \ub g_{5143} &  \ub g_{5243} \\[0.2cm]
 0 & - \ub g_{1435} & 0 & \ub g_{4213} & - \ub g_{4132}& 0&-\ua g_{4513} & 0 &  \ub g_{1543} & 0 \\[0.2cm]

  \ub g_{1254} &  \ub g_{1354} & 0 & 0 & 0 & 0& - \ub g_{5143} & - \ub g_{1543}&0& \ua g_{5124} \\[0.2cm]
 \ub g_{2154} & 0 & 0 & - \ub g_{2541} & 0 & \ub g_{5324} & - \ub g_{5243}& 0&-\ua g_{5124} & 0 

\ea
\right),}
\ee
where:
\be\label{dirac11}
\ba{rcl}
\ua g_{ijkl} &=&- \dfrac{1}{\gamma}  \dfrac{1}{A_{ik+}} \csc\phi_{ijk+}\csc\phi_{ilk+} \sin\alpha_{ik;jl+} ,\\[0.5cm]
\ub g_{ijkl} &=&- \dfrac{1}{\gamma} \dfrac{1}{A_{ij+}} \cot\phi_{ijk+}\csc\phi_{ilj+} \sin\alpha_{ij;lk+}.
\ea
\ee
In choosing the constraints as we did, we have postponed the pain to the following evaluation:
\be\label{dirac09}
\{\scf, \theta_+\}_L =  \{\phi_+, \theta_+\}_L - \frac{\partial \Phi_+(A)}{\partial A_+} \{A_+, \theta_+\}_L\; ,
\ee
While we cannot specify $\Phi_+(A)$ explicitly, we can in principle circumnavigate this obstacle for the 4-simplex boundary since the 10 areas $\{A_+\}$ actually already specify a set of 10 edge lengths $\{l_+\}$ for the boundary in question.  Let us remark that these lengths $\{l_+\}$ are not identical to the $\{L_+\}$ defined in Appendix \ref{app:gauge-defs} at this stage, although they do coincide on the geometrical phase space.   Moreover, the $\Phi_+$ may be written as explicit functions of these edge lengths, that is, as $\Phi_+(A(l))$.   This manipulation allows us to calculate:
\be
\frac{\partial \Phi_+(A)}{\partial A_+} = \frac{\partial \Phi_+(A(l))}{\partial l_+} \frac{\partial l_+}{\partial A_+} = \frac{\partial \Phi_+(A(l))}{\partial l_+} \left(\frac{\partial A_+}{\partial l_+}\right)^{-1}
\ee
To clarify, for a 4-simplex boundary:
\be\label{spec02}
\ba{rcl}
A_{ij}(l) &=& \dfrac{1}{4}\left[ 2l^2_{ijk}l^2_{ijl} + 2l^2_{ijl}l^2_{ijm} + 2l^2_{ijm}l^2_{ijk} - l_{ijk}^4 - l_{ijl}^4 - l_{ijm}^4  \right]^{\frac{1}{2}},\\ [0.4cm]
\cos\hat\Phi_{ijk}(l) &=& \dfrac{1}{8A_{ij}A_{ik}}\Big( l^2_{ijk} (l^2_{ilj} + l^2_{ilk} - l^2_{ilm}) - \dfrac{1}{2} (l^2_{ijk} + l^2_{ijl} - l^2_{ijm})(l^2_{ikj}  + l^2_{ikl} - l^2_{ikm})\Big), \\[0.4cm]

\cos\hat\alpha_{ij,kl} &=& \dfrac{1}{2l_{ijk}l_{ijl}}\Big( l^2_{ijk} + l^2_{ijl} - l^2_{ijm}\Big), \quad \quad\quad \quad \sin\hat\alpha_{ij,kl} = 2\dfrac{A_{ij}}{l_{ijk}l_{ijl}}, 
\ea
\ee
where the hats record the fact that in the main text we were dealing with exterior dihedral angles, while here we are defining interior dihedral angles. Thus, $\hat\Phi = \pi - \Phi$ and $\hat\alpha = \pi - \alpha$.  Under this transformation, there are a number of important sign changes.     Additionally, the subscripts $ijklm  = \sigma(12345)$ where $\sigma \in S_5$.   The relevant derivatives are straightforward to calculate:
\be\label{spec03}
{\renewcommand{\arraystretch}{3}
\ba{rcl}
\dfrac{\partial A_{ij}(l)}{\partial l_{ijk}} &=& \dfrac{l_{ijk}}{8A_{ij}}\Big(l^2_{ijl} + l^2_{ijm} - l^2_{ijk} \Big) = \dfrac{1}{2}l_{ijk}\cot\hat\alpha_{ij;lm}\;,\\

\dfrac{\partial \cos\hat\Phi_{ijk}(l)}{\partial l_{ijk}} &=&\dfrac{l_{ijk}}{2}\left(\dfrac{ \cot \hat\alpha_{ij;lm}}{A_{ik}} + \dfrac{\cot \hat\alpha_{ij;ml}}{A_{ij}}\right)   - \dfrac{l_{ijk}}{2}\left(\dfrac{ \cot \hat\alpha_{ij;lm}}{A_{ij}} + \dfrac{\cot \hat\alpha_{ij;ml}}{A_{ik}}\right) \cos\hat\Phi_{ijk}  - \dfrac{l_{ijk}l_{ilm}^2}{4A_{ij}A_{ik}}\\

\dfrac{\partial \cos\hat\Phi_{ijk}(l)}{\partial l_{ijl}} &=& \dfrac{l_{ijl}}{2A_{ik}}\cot\hat\alpha_{ik;jm} - \dfrac{l_{ijl}}{2A_{ij}} \cos\hat\Phi_{ijk}\cot\hat\alpha_{ij;mk},\\

\dfrac{\partial \cos\hat\Phi_{ijk}(l)}{\partial l_{ilm}} &=& \dfrac{1}{4} \dfrac{l_{ijk}^2 l_{ilm}}{A_{ij}A_{ik}} ,
\ea}
\ee
While it may seem plain sailing from now on, one should remember that we have yet to invert the matrix $\frac{\partial A}{\partial l}$.  While some progress can be made for general configurations, we must admit that we have not found a way to simplify the sums of products of cotangents.  Thus, we specialize yet more to the equilateral configuration in order to perform an explicit inversion.

%



%



%

%

%

\subsection{The equilateral 4-simplex}
\label{spec}
 Now, we must invert the matrix $\dfrac{\partial A(l)}{\partial l}$ to get the matrix of partial derivatives: $\dfrac{\partial l(A)}{\partial A}$.  We shall order the areas and lengths as:
\be\label{spec04}
\ba{llllllllll}
A_{12}&A_{13}&A_{23}&A_{24}&A_{34}&A_{35}&A_{45}&A_{41}&A_{51}&A_{52},\\ [0.1cm]
l_{125}&l_{135}&l_{231}&l_{241}&l_{342}&l_{352}&l_{453}&l_{413}&l_{514}&l_{524}.
\ea
\ee
Furthermore, we shall consider only the equilateral case in which:
\be\label{spec05}
\ba{rclcrcl}
A & = & \dfrac{\sqrt{3}}{4}l^2,\\[0.2cm]
\cos\hat\alpha & = & \dfrac{1}{2}, && \sin\hat\alpha & = & \dfrac{\sqrt{3}}{2},\\[0.2cm]
\cos\hat\phi & = & \dfrac{1}{3}, && \sin\hat\phi & = & \dfrac{2\sqrt{2}}{3},\\[0.2cm]
\cos\hat\theta & = & \dfrac{1}{4}, && \sin\hat\theta & = & \dfrac{\sqrt{15}}{4}.\\[0.2cm]
\ea
\ee
Then, the matrix evaluations yield:
\be\label{spec06} 
  \dfrac{\partial A(l)}{\partial l} = \dfrac{l}{2\sqrt{3}}
\left(
{\scriptsize\addtolength{\arraycolsep}{-0.5mm}
\ba{cccccccccc}
1 & 0 & 1 & 1 & 0 & 0 & 0 & 0 & 0 & 0\\[0.2cm]

0 & 1 & 1 & 0 & 0 & 0 & 0 & 1 & 0 & 0\\[0.2cm]

0 & 0 & 1 & 0 & 1 & 1 & 0 & 0 & 0 & 0\\[0.2cm]

0 & 0 & 0 & 1 & 1 & 0 & 0 & 0 & 0 & 1\\[0.2cm]

0 & 0 & 0 & 0 & 1 & 0 & 1 & 1 & 0 & 0\\[0.2cm]

0 & 1 & 0 & 0 & 0 & 1 & 1 & 0 & 0 & 0\\[0.2cm]

0 & 0 & 0 & 0 & 0 & 0 & 1 & 0 & 1 & 1\\[0.2cm]

0 & 0 & 0 & 1 & 0 & 0 & 0 & 1 & 1 & 0\\[0.2cm]

1 & 1 & 0 & 0 & 0 & 0 & 0 & 0 & 1 & 0\\[0.2cm]

1 & 0 & 0 & 0 & 0 & 1 & 0 & 0 & 0 & 1

\ea}
\right),
\quad
\dfrac{\partial l(A)}{\partial A} = \dfrac{1}{\sqrt{3}l}
\left(
{\scriptsize\addtolength{\arraycolsep}{-1mm}
\ba{rrrrrrrrrr}

2 & -1 & -1 & -1 & 2 & -1 & -1 & -1 & 2 & 2\\[0.2cm]

-1 & 2 & -1 & 2 & -1 & 2 & -1 & -1 &  2 & -1\\[0.2cm]

2 & 2 & 2 & -1 & -1 & -1 & 2 & -1 & -1 & -1\\[0.2cm]

2 & -1 & -1 & 2 & -1 & 2 & -1 &  2 & -1 & -1 \\[0.2cm]

-1 & -1 & 2 & 2 & 2 & -1 & -1 & -1 & 2 & -1 \\[0.2cm]

-1 & -1 & 2 & -1 & -1 & 2 & -1 & 2 & -1 & 2 \\[0.2cm]

2 & -1 & -1 & -1 & 2 & 2 & 2 & -1 & -1 & -1 \\[0.2cm]

-1 & 2 & -1 & -1 & 2 & -1 & -1 & 2 & -1 & 2 \\[0.2cm]

-1 & -1 & 2 & -1 & -1 & -1 & 2 & 2 & 2 & -1 \\[0.2cm]

-1 & 2 & -1 & 2 & -1 & -1 & 2 & -1 & -1 & 2

\ea}
\right),\nonumber
\ee
\be
\dfrac{\partial \cos\hat\Phi(l)}{\partial l} = 
 \dfrac{4}{9l}
\left(
{\scriptsize\addtolength{\arraycolsep}{-1.4mm}
\ba{rrrrrrrrrr}

-1 & 1 & 1 & 1 & 0 & 0 & 0 & -3 & 1 & 0\\[0.2cm]

1 & -1 & 1 & -3 & 0 & 0 & 0 & 1 &  1 & 0\\[0.2cm]

1 & 0 & -1 & 1 & 1 & 1 & 0 & 0 & 0 & -3\\[0.2cm]

1 & 0 & 1 & -1 & 1 & -3 & 0 &  0 & 0 & 1 \\[0.2cm]

0 & -3 & 1 & 0 & -1 & 1 & 1 & 1 & 0 & 0 \\[0.2cm]

0 & 1 & 1 & 0 & 1 & -1 & 1 & 3 & 0 & 0 \\[0.2cm]

0 & 0 & 0 & -3 & 1 & 0 & -1 & 1 & 1 & 1 \\[0.2cm]

0 & 0 & 0 & 1 & 1 & 0 & 1 & -1 & 1 & 3 \\[0.2cm]

1 & 1 & 0 & 0 & 0 & -3 & 1 & 0 & -1 & 1 \\[0.2cm]

1 & -3 & 0 & 0 & 0 & 1 & 1 & 0 & 1 & -1

\ea}
\right)
%
%
%
%
%
%
%
%
%
%
%
%
%
%
\;.\nonumber
\ee
Furthermore: 
\be\label{spec09}
 \{A_+, \theta_+\}_L = 
\left(
{\scriptsize\addtolength{\arraycolsep}{-0.5mm}
\ba{rrrrrrrrrr}
1 & 0 & 0 & 0 & 0 & 0 & 0 & 0 & 0 & 0 \\[0.2cm]

0 & 1 & 0 & 0 & 0 & 0 & 0 & 0 & 0 & 0 \\[0.2cm]

0 & 0 & 1 & 0 & 0 & 0 & 0 & 0 & 0 & 0 \\[0.2cm]

0 & 0 & 0 & 1 & 0 & 0 & 0 & 0 & 0 & 0 \\[0.2cm]

0 & 0 & 0 & 0 & 1 & 0 & 0 & 0 & 0 & 0 \\[0.2cm]

0 & 0 & 0 & 0 & 0 & 1 & 0 & 0 & 0 & 0 \\[0.2cm]

0 & 0 & 0 & 0 & 0 & 0 & 1 & 0 & 0 & 0 \\[0.2cm]

0 & 0 & 0 & 0 & 0 & 0 & 0 & 1 & 0 & 0 \\[0.2cm]

0 & 0 & 0 & 0 & 0 & 0 & 0 & 0 & 1 & 0 \\[0.2cm]

0 & 0 & 0 & 0 & 0 & 0 & 0 & 0 & 0 & 1 
\ea}
\right),
\quad\quad
\dr^{\!\!\!\!-1} = 
\dfrac{l^2}{2\gamma}
\left(
{\scriptsize\addtolength{\arraycolsep}{-0.7mm}
\ba{rrrrrrrrrr}
0 & -1 & 0 & 0 & 0 & 0 & 0 & 0 & 0 & 0 \\[0.2cm]

\phantom{-}1 & 0 & 0 & 0 & 0 & 0 & 0 & 0 & 0 & 0 \\[0.2cm]

0 & 0 & 0 & -1 & 0 & 0 & 0 & 0 & 0 & 0 \\[0.2cm]

0 & 0 & \phantom{-}1 & 0 & 0 & 0 & 0 & 0 & 0 & 0 \\[0.2cm]

0 & 0 & 0 & 0 & 0 & -1 & 0 & 0 & 0 & 0 \\[0.2cm]

0 & 0 & 0 & 0 & \phantom{-}1 & 0 & 0 & 0 & 0 & 0 \\[0.2cm]

0 & 0 & 0 & 0 & 0 & 0 & 0 & -1 & 0 & 0 \\[0.2cm]

0 & 0 & 0 & 0 & 0 & 0 & \phantom{-}1 & 0 & 0 & 0 \\[0.2cm]

0 & 0 & 0 & 0 & 0 & 0 & 0 & 0 & 0 & -1 \\[0.2cm]

0 & 0 & 0 & 0 & 0 & 0 & 0 & 0 & \phantom{-}1 & 0 
\ea}
\right),
\ee
while:\footnote{There is a subtlety, which we have taken care of,  in the signs of some of the terms.  $V_{ijkl}$ is  the volume squared up to a sign depending on the permutation of the indices.  It is important to chose one to be positive for each tetrahedron.}
\be\label{spec10}
\{\phi_+,\theta_+\}_L = 
\dfrac{1}{l^2}\sqrt{\dfrac{1}{6}}
\left(
{\scriptsize\addtolength{\arraycolsep}{-1.5mm}
\ba{rrrrrrrrrr}
4 & -3 & 0 & 0 & 0 & 0 & 0 & 0 & 1 & 0\\[0.2cm]
-3& 4 & 0 & 0 & 0 & 0 & 0 & -3 & 1 & 0\\[0.2cm]

1 & 0& 4 & -3 & 0 & 0 & 0 & 0 & 0 & 0\\[0.2cm]
 1 & 0&- 3& 4 & 0 & 0 & 0 & 0 & 0 & -3\\[0.2cm]

0 & 0& 1 & 0& 4 & -3 & 0 & 0 & 0 & 0  \\[0.2cm]
0 & -3 & 1 & 0& -3 & 4 & 0 & 0 & 0 & 0  \\[0.2cm]

0 & 0 & 0 & 0& 1 & 0& 4 & -3 & 0 & 0  \\[0.2cm]
0 & 0 & 0 & -3 & 1 & 0& -3& 4 & 0 & 0  \\[0.2cm]

0 & 0 & 0 & 0 & 0 & 0& 1 & 0& 4 & -3  \\[0.2cm]
 0 & 0 & 0 & 0 & 0 & -3 & 1 & 0& -3& 4 \ea}
\right),\quad
\{\theta_+,\theta_+\}_L = 
 \dfrac{3}{4\gamma l^2}
\left(
{\scriptsize\addtolength{\arraycolsep}{-1.5mm}
\ba{rrrrrrrrrr}
0 & 3 & -1 & 1 & 0 & 0 & 0 & 0 & 1 & -1\\[0.2cm] 
-3 & 0 & 1 & 0 & 0 & -1 & 0 & 1 & -1 & 0\\[0.2cm] 
1 & -1 & 0 & 3 & -1 & 1 & 0 & 0 & 0 & 0\\[0.2cm] 
-1 & 0 & -3 & 0 & 1 & 0 & 0 & -1 & 0 & 1\\[0.2cm] 
0 & 0 & 1 & -1 & 0 & 3 & -1 & 1 & 0 & 0\\[0.2cm] 
0 & 1 & -1 & 0 & -3 & 0 & 1 & 0 & 0 & -1\\[0.2cm] 
0 & 0 & 0 & 0 & 1 & -1 & 0 & 3 & -1 & 1\\[0.2cm] 
0 & -1 & 0 & 1 & -1 & 0 & -3 & 0 & 1 & 0\\[0.2cm] 
-1 & 1 & 0 & 0 & 0 & 0 & 1 & -1 & 0 & 3\\[0.2cm] 
1 & 0 & 0 & -1 & 0 & 1 & -1 & 0 & -3 & 0
\ea}
\right),
\ee
Ultimately, one finds:
\be\label{spec12}
\{\theta_{+}, \theta_{+}\}_R = \{\theta_{+}, \theta_{+}\}_L - \{\theta_+,\scf\}_L  \dri \{\scf, \theta_+\}_L    = 0,
\ee
which is the result we were expecting.




\vspace{0.5cm}

~\\
{\bf \large Acknowledgements}
~\\

We thank Sergei Alexandrov, Daniele Oriti, Roberto Pereira and Simone Speziale for discussions. Research at Perimeter Institute is supported by the Government of Canada through Industry Canada and by the Province of Ontario through the Ministry of Research and Innovation.

{\footnotesize

\bibliographystyle{utphys}

\bibliography{papers}

}

\end{document}